\documentclass[preprint,12pt,3p]{elsarticle}

\usepackage{amsmath,amssymb,amsfonts}
\usepackage{graphicx}
\usepackage{booktabs}
\usepackage{multirow}
\usepackage{xcolor}
\usepackage{hyperref}
\usepackage{lineno}
\usepackage{subcaption}
\usepackage{natbib}
\usepackage{threeparttable}
\biboptions{sort&compress}
\usepackage{comment}
\usepackage{tikz}
\usepackage{url}
\usetikzlibrary{positioning}

\journal{Chaos, Solitons \& Fractals}

\begin{document}
\begin{frontmatter}

\title{Persistence without Multifractality in Tropical Atlantic SST Indices: Scaling-Range Artefacts and a Reappraisal of the ENSO Association}

\author[1]{Sebasti\'an Jaroszewicz}
\author[2]{Nahuel Mendez\corref{cor1}}
\ead{nahueldanielmendez@gmail.com}
\author[3]{Maria P. Beccar-Varela}
\author[3]{Maria Cristina Mariani}

\address[1]{Comisi\'on Nacional de Energ\'ia At\'omica, Buenos Aires, Argentina}
\address[2]{Instituto S\'abato, Buenos Aires, Argentina}
\address[3]{Department of Mathematical Sciences, University of Texas at El Paso, El Paso, United States}

\cortext[cor1]{Corresponding author}

\begin{abstract}
We investigate the scaling properties of three weekly sea-surface-temperature (SST) indices of the
tropical Atlantic---two regional indices (SAT, TSA) and the interhemispheric gradient (TASI)---over
1981--2025 using multifractal detrended fluctuation analysis (MFDFA) and its sign-preserving
cross-correlation extension (MFCCA). Exponents are estimated over $s\in[18,100]$ weeks: the lower
bound satisfies the condition $s_{\min}\gtrsim6(m+1)$ imposed by the second-order detrending adopted
here, and the upper bound is strictly constrained to precede the scaling crossover observed
beyond $\approx100$ weeks. On these scales all three indices are strongly persistent and locally
non-stationary, with $h(2)=1.32$--$1.35$ exceeding unity at every detrending order tested. Testing the apparent multifractal widths against ensembles
of shuffled and phase-randomised (IAAFT) surrogates, we find that none of the three is multifractal
within the resolution afforded by the record. The widths of SAT and TASI ($\Delta h=0.077$ and
$0.054$) are fully compatible with both nulls. The width of TSA ($\Delta h=0.217$) exceeds both, but
is excluded on structural rather than statistical grounds: its generalized Hurst exponent increases
with $q$, and since $f(\alpha)=1+q^{2}h'(q)$, an increasing $h(q)$ forces $f(\alpha)>1$, a value no
one-dimensional record can attain. Its apparent width is therefore a signature of a single power law
fitted across a change of scaling regime rather than of a multiplicative cascade, and the singularity
spectrum is accordingly reported as a consistency check on $h(q)$ rather than as a result. The apparent widths are in any case not identifiable at this record length: with $N\approx2300$ and about
three-quarters of a decade of usable scales, each of the three indices is the widest of the three under
some defensible combination of fitting range, moment range and detrending order, and their bootstrap
confidence intervals overlap throughout. The increments are only mildly
leptokurtic, with excess kurtosis below unity, so the extreme moments are not populated by heavy
tails. The pairwise couplings are strong--the regional indices co-vary positively, increasingly so
towards longer scales, whereas the gradient is anticorrelated with both--they are governed by a single cross-scaling exponent, confirming their monofractal nature. A lagged analysis further shows that a previously reported association
between the gradient and ENSO does not survive a surrogate test accounting for the multiplicity of
tested lags, being consistent with the shared persistence of the two series. These results revise
earlier estimates of multifractal strength in tropical Atlantic SST substantially downward, and
indicate that reported widths should not be interpreted before the sign of $h'(q)$ and the stability
of the fitting range have been verified.
\end{abstract}


\end{frontmatter}

\section{Introduction}

Sea surface temperature (SST) variability in the Tropical Atlantic constitutes a key node in the global climate system, modulating regional precipitation patterns, atmospheric circulation, and inter-basin teleconnections \cite{kushnir2010mechanisms,huo2015role,hurrell1999global}. To characterize this variability, the oceanographic and climate communities have developed a set of diagnostic indices — among them the South Atlantic Tropical (SAT) index, the Tropical Southern Atlantic (TSA) index, and the interhemispheric SST gradient (TASI), which captures the cross-equatorial thermal contrast between the Tropical North and South Atlantic \cite{chang1997decadal}. These indices have been extensively studied using linear statistical frameworks, including empirical orthogonal functions, spectral analysis \cite{andreoli2004multi}, and Pearson correlation \cite{prestes2025climate}, which have established their connections to regional rainfall anomalies \cite{andreoli2006tropical}, Atlantic hurricane activity \cite{trenberth2006atlantic}, and the El Ni\~no–Southern Oscillation (ENSO) \cite{huang2005atlantic,venegas1997atmosphere,trenberth1997definition}.

However, there is growing evidence that oceanic and atmospheric variability exhibits pronounced nonlinearity, long-range temporal correlations, and scale-dependent intermittency — properties that are structurally incompatible with linear or stationary analysis frameworks \cite{robinson2003dynamical,dippe2018relationship}. Tropical ocean–atmosphere interactions, in particular, involve multiplicative cascade processes, threshold-driven regime shifts, and cross-scale energy transfers that can only be characterized within a nonlinear scaling paradigm. These features motivate the application of multifractal analysis to climate time series, as a means of quantifying the full hierarchy of scaling exponents that characterize such systems beyond the single-exponent description afforded by classical Hurst analysis \cite{reljin2000multifractal,baranowski2015multifractal}.

Multifractal Detrended Fluctuation Analysis (MFDFA), introduced by Kantelhardt et al.\cite{Kantelhardt_2002}, has emerged as the standard method for quantifying multifractal properties in non-stationary geophysical time series. By computing the generalized Hurst exponent \(h(q)\) across a range of moment orders \(q\), MFDFA provides a complete characterization of the singularity spectrum \(f(\alpha)\), whose width \(\Delta h\) measures the degree of multiscale heterogeneity in the signal \cite{ihlen2012introduction}. The method has been applied to a wide range of climate records, including sea surface temperatures \cite{lim2024auto}, precipitation series \cite{lim2024auto}, atmospheric pressure fields, and coupled ocean–atmosphere systems \cite{fuwape2023multifractal}. Crucially, when combined with surrogate data testing — in particular the Iterated Amplitude-Adjusted Fourier Transform (IAAFT) algorithm \cite{schreiber2000surrogate, schreiber1996improved}— MFDFA allows the physical sources of multifractality to be disentangled: a significant reduction in \(\Delta h\) under phase randomization uniquely identifies a contribution from nonlinear phase correlations, as distinct from linear persistence. Temporal correlations are a necessary condition for genuine multifractality, and a broad distribution of fluctuations can widen the singularity spectrum only when such correlations are already present~\cite{Kwapien2023,Rak2025}; the two are therefore not independent
sources but a hierarchy, and the surrogate tests below are interpreted accordingly \cite{mendez2026characterising}.

Despite these advances, the comparative multifractal structure of Tropical Atlantic SST indices has not been systematically investigated. Previous studies have applied MFDFA to individual regional indices or gridded SST fields \cite{bunimovich1993observations}, but a comparative analysis that distinguishes between regional indices (SAT, TSA) and the interhemispheric gradient (TASI) — which integrates variability across both hemispheres and captures cross-equatorial wind–SST feedbacks — remains absent from the literature. This distinction is physically significant: while SAT and TSA reflect relatively localized thermodynamic adjustments, the TASI gradient encodes the full meridional reorganization of the tropical Atlantic, including the Atlantic Meridional Mode (AMM) and cross-equatorial interactions that are known to produce genuinely nonlinear dynamics \cite{vimont2007atlantic}.

A second important limitation of existing multifractal studies of climate is their predominantly static character. Most analyses characterize multifractal properties over a full observational record, overlooking the temporal evolution of scaling behavior, which may carry important information about regime transitions and externally forced variability. This limitation is particularly relevant in the context of inter-basin teleconnections: if ENSO exerts a modulating influence on Tropical Atlantic complexity, such an influence has been hypothesized to be episodic and threshold-dependent, manifesting primarily during extreme events rather than as a persistent background coupling. A dynamic, time-resolved approach, tested against appropriate surrogates, is therefore necessary to determine whether such signatures are present.

In this work, we address both gaps by means of a multi-component framework. We perform a comparative global MFDFA of SAT, TSA, and TASI, including surrogate data testing to disentangle the sources of multifractality. We then introduce a moving-window analysis that resolves the scaling exponents in time, focusing on the local Hurst exponent \(h_2(t)\) and testing the local multifractal width against window-specific surrogates. Finally, we test the statistical association between the Atlantic gradient (TASI) and ENSO variability — as represented by the Oceanic Ni\~no Index (ONI) — using a lagged detrended cross-correlation analysis with a family-wise surrogate null that accounts for the search over lags.

The paper is organized as follows. Section 2 describes the data sources and preprocessing. Section 3 presents the methodological framework. Section 4 reports and discusses the results. Section 5 states the conclusions and identifies directions for future work.

\section{Data}
\label{sec:data}

\begin{figure}
    \centering
    \includegraphics[width=1\linewidth]{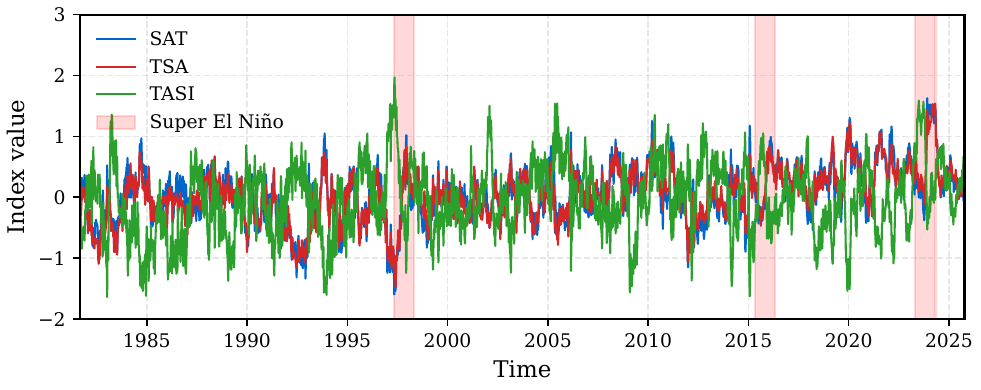}
    \caption{Weekly sea surface temperature (SST) anomalies for the Tropical Atlantic indices SAT, TSA, and TASI over the period 1981–2025. Shaded regions indicate major El Niño events (1997–1998 and 2015–2016). The TASI index, defined as the interhemispheric SST gradient, exhibits enhanced variability compared to the regional indices, reflecting the combined influence of both hemispheres.}
    \label{fig:indices}
\end{figure}

The primary dataset for the Tropical Atlantic consists of the South Atlantic Tropical (SAT) index and the Tropical Southern Atlantic (TSA) index, both obtained from the NOAA Physical Sciences Laboratory (PSL) \citep{PSL_SAT, PSL_TSA}. The inter-hemispheric Tropical Atlantic SST Index (TASI) was computed following the definitions commonly established in the tropical Atlantic literature \citep{Enfield_1999, Enfield_2001}.
The inter-hemispheric Tropical Atlantic SST Index (TASI) was defined in \cite{Chang_1997} and is formally calculated as the difference in sea surface temperature anomalies between the Tropical North Atlantic (TNA) and the Tropical South Atlantic (TSA). Geographically, it is calculated as:
$$\text{TASI}=\text{SST}_\text{TNA}-\text{SST}_\text{TSA}$$

where the TNA region is typically bounded by 5$^{\circ}$N--20$^{\circ}$N,
60$^{\circ}$W--30$^{\circ}$W, and the TSA region by 0$^{\circ}$--20$^{\circ}$S,
30$^{\circ}$W--10$^{\circ}$E \cite{Enfield_1999}.
These Atlantic records consist of high-frequency weekly SST anomalies spanning
from September 1981 to October 2025, yielding approximately 2,300 observations
per series (Fig.~\ref{fig:indices}). Missing values within the weekly records
were exceptionally sparse and were handled via linear interpolation. All
Atlantic indices are expressed in degrees Celsius (anomaly units).

To quantify the external global forcing and assess the non-linear inter-oceanic teleconnections, we incorporated the Oceanic Ni\~no Index (ONI) as the standard indicator of the El Ni\~no-Southern Oscillation (ENSO). The ONI data, which represents the 3-month running mean of SST anomalies in the Ni\~no 3.4 region (5$^\circ$N-5$^\circ$S, 120$^\circ$-170$^\circ$W), were retrieved from the NOAA Climate Prediction Center (CPC). 

For the univariate global and moving-window MFDFA assessments, the high-frequency weekly resolution of the Atlantic indices was preserved to capture sub-seasonal multifractal dynamics. However, for the bivariate coupling analysis between the Atlantic gradient and ENSO, the weekly Atlantic series were resampled to a monthly frequency (via monthly averaging) to match the temporal resolution of the ONI.

\section{Methodology}
\label{sec:methods}

\subsection{Multifractal Detrended Fluctuation Analysis (MFDFA)}
\label{sec_mfdfa}

To characterize the scale-dependent complexity of the SST indices, we applied the Multifractal Detrended Fluctuation Analysis (MFDFA) \cite{Kantelhardt_2002}. For each discrete time series $x(i)$ of length $N$, we first construct the integrated profile:
\begin{equation}
Y(i) = \sum_{k=1}^{i} [x(k) - \langle x \rangle]
\end{equation}
where $\langle x \rangle$ is the mean of the series. The profile $Y(i)$ is divided into $N_s \equiv \lfloor N/s \rfloor$ non-overlapping segments of scale $s$. To account for data length that is not a multiple of $s$, the procedure is repeated from the opposite end of the series, yielding $2N_s$ segments. Within each segment $v$, a local polynomial trend of order $m$ is fitted and subtracted; we adopt a second-order detrending ($m=2$) as the primary analysis, for the reasons discussed in Section~\ref{sec:origin}. The detrended variance is then calculated, and the $q$-th order fluctuation function is obtained by averaging across all segments:
\begin{equation}
F_{q}(s) = \left\{ \frac{1}{2N_{s}} \sum_{v=1}^{2N_{s}} [F^2(s,v)]^{\frac{q}{2}} \right\}^{\frac{1}{q}}
\label{eq:Fq_univar}
\end{equation}
where $q$ is the moment order. For $q=0$, a logarithmic averaging procedure is used as the analytical limit. If the series exhibits scale-invariant properties, the fluctuation function scales as a power law:
\begin{equation}
F_{q}(s) \sim s^{h(q)}
\end{equation}
where $h(q)$ is the generalized Hurst exponent. The multifractal scaling exponent (mass exponent) $\tau(q)$ is directly related to $h(q)$ by $\tau(q) = q h(q) - 1$. The singularity spectrum $f(\alpha)$, which describes the fractal dimension of the subset of series characterized by the singularity strength (H\"older exponent) $\alpha$, is obtained via a Legendre transform: $\alpha = d\tau/dq$ and $f(\alpha) = q\alpha - \tau(q)$.
Combining these two relations with $\tau=qh-1$ gives $f(\alpha)=1+q^{2}h'(q)$, so that
$f(\alpha)\le1$---the bound imposed by the dimension of the support of a one-dimensional
record---holds if and only if $h(q)$ is non-increasing. We use this as a consistency check on the
estimated $h(q)$ in Section~\ref{sec:global}.

In this study, we evaluated the scaling behavior across scales $s \in [18, 100]$ weeks, using moments
$q \in [-4, 4]$ for the primary analysis; the sensitivity of the estimated widths to a wider moment
range is examined in Section~\ref{sec:origin}. The minimum scale is set by the detrending order: a
polynomial of order $m$ is not determined within a segment appreciably shorter than $6(m+1)$ points,
which gives $s_{\min}=18$ for the second-order detrending adopted here, and we apply the same rule
whenever the order is varied. The maximum scale is set by the data rather than by a fixed fraction of
the record length: the fluctuation functions scale cleanly up to $s\approx100$ weeks and change slope
beyond it (Supplementary~S1), so that a single power law fitted over the full accessible interval
would mix two regimes. Over the fitted range the log--log plots of $\log_{10}F_q(s)$ against
$\log_{10}s$ are linear for every index and every moment of the primary range, with coefficients of
determination $R^{2}\ge0.975$ throughout (minima $0.975$, $0.991$ and $0.992$ for SAT, TSA and TASI
respectively). Both the fluctuation functions and the moment-by-moment fit quality are reported in
Supplementary~S1, together with the fitted range and its complement, so that the basis for the choice
can be inspected directly.

To rigorously estimate the statistical uncertainties of the multifractal spectra, we implemented a residual bootstrapping technique on the Ordinary Least Squares (OLS) regression of the fluctuation functions. For each moment $q$, the residuals of the $\log_{10} F_q(s)$ versus $\log_{10} s$ linear fit were resampled with replacement over $N = 2000$ iterations. This procedure allows us to construct the empirical probability distribution of the regression slopes (the generalized Hurst exponents, $h(q)$) and properfly propagate the standard error to the multifractal width ($\Delta h$). Throughout this study, multifractal parameters are reported with their associated 95\% confidence intervals.

\subsection{Surrogate Data Testing: Shuffled and IAAFT Series}
To definitively disentangle the physical sources of the observed multifractality, we tested the original series against two distinct null hypotheses using surrogate data generation. First, we generated \textit{randomly shuffled} series, which preserve the empirical probability density function (PDF) and extreme values (heavy tails) of the original data but completely destroy temporal correlations. A collapse of the multifractal width ($\Delta h_{shuf} \to 0$, $h(q) \to 0.5$) indicates that the multifractality is fundamentally driven by temporal memory rather than the amplitude distribution.

It should be emphasized that these two channels are not independent sources
acting in parallel. Genuine multifractality requires temporal correlations: in
their absence the singularity spectrum of a finite record is
non-degenerate only through finite-size effects, which vanish asymptotically at
positive moments as $N$ grows. A broad distribution of fluctuations does not by
itself generate multifractality; it can only broaden a spectrum that
correlations have already produced~\cite{Kwapien2023,Rak2025}. The shuffled
ensemble therefore does not isolate a ``distributional multifractality'' but
rather measures the finite-size floor against which any genuine width must be
judged.

Second, to separate linear persistence from non-linear phase coupling, we utilized the Iterated Amplitude Adjusted Fourier Transform (IAAFT) algorithm \cite{schreiber1996improved}. The IAAFT surrogates strictly preserve both the original amplitude distribution and the linear auto-correlation structure (the power spectrum) but randomize the Fourier phases. Therefore, a significant reduction in multifractality in the IAAFT surrogates isolates the presence of non-linear interactions and multiplicative cascading processes.

\subsection{Non-Stationary Complexity: Moving-Window MFDFA}
\label{sec:moving_window}
To assess the non-stationarity of the Atlantic thermodynamic regimes, we extended the static MFDFA framework to a dynamic, time-resolved approach. The MFDFA was computed over sliding windows of 5 years (260 weeks), shifted forward with a step size of 3 months (13 weeks). Because a window of this length does not support a reliable estimate of the full multifractal spectrum, we take the local Hurst exponent $h_2(t)$---the exponent at $q=2$, which is robustly estimable at this window length---as the primary time-resolved measure. Where a local multifractal width $\Delta h(t)$ is reported, it is assessed only in conjunction with a window-specific significance test: for each window we generated $K=50$ IAAFT surrogates of the enclosed segment, from which we obtained the expected width, its standard deviation, and a one-sided per-window $p$-value. Robustness to the window length was checked with windows of 3, 5 and 7 years.

To ensure robust power-law estimation within each local block, the scaling range for the
moving-window analysis was adapted to the window length ($N_{\text{local}} = 260$ weeks), with
$s \in [10, 65]$ weeks, so that $s_{\text{max}} \le N_{\text{local}}/4$ and at least four
non-overlapping segments are available at the largest scale. All windowed quantities use second-order
detrending, consistent with the global analysis. The lower bound here is below the value
$s_{\min}=18$ imposed by the detrending order in the global analysis, which the window length does
not allow us to raise without leaving too short a fitting interval; the windowed results are
therefore used only for the local Hurst exponent $h_2(t)$, whose diagnosis of persistence is
qualitative and robust, and not for any statement about local multifractal width.

\subsection{Cross-correlation analysis and the ENSO association}
\label{sec:entropy_and_mfdcca}
To quantify the coupling among the indices, and between the Atlantic gradient and ENSO, we used the sign-preserving multifractal cross-correlation analysis (MFCCA)~\cite{Oswiecimka2014}, in which the detrended covariance retains its sign. Both series are integrated into profiles $Y_x(i)$ and $Y_y(i)$ as in Section~\ref{sec_mfdfa} and partitioned into the same $2N_s$ segments of scale $s$. Within each segment $\nu$ a polynomial trend of order $m=2$ is fitted to each profile separately and the \emph{signed} detrended cross-covariance is computed as
\begin{equation}
F^{2}_{xy}(s,\nu)=\frac{1}{s}\sum_{k=1}^{s}
\Big[Y_x\big(\nu_{k}\big)-\widetilde{Y}_x^{\,\nu}\big(\nu_{k}\big)\Big]
\Big[Y_y\big(\nu_{k}\big)-\widetilde{Y}_y^{\,\nu}\big(\nu_{k}\big)\Big],
\end{equation}
where $\widetilde{Y}^{\,\nu}$ denotes the local polynomial fit within segment $\nu$ and $\nu_k$ indexes the points of that segment Unlike the univariate case, $F^{2}_{xy}(s,\nu)$ is a covariance and may be negative. Following Oświęcimka et al.~\cite{Oswiecimka2014}, the sign is retained and carried into the generalized moments, which is what distinguishes MFCCA from the earlier MF-DCCA prescription based on $|F^{2}_{xy}|$. Together with the corresponding variances $F^{2}_{zz}(s,\nu)$ of each detrended profile ($z$ standing for either $x$ or $y$), this defines the family of $q$-th order fluctuation functions
\begin{align}
F^{q}_{xy}(s)&=\frac{1}{2N_{s}}\sum_{\nu=1}^{2N_{s}}
  \mathrm{sign}\!\left[F^{2}_{xy}(s,\nu)\right]
  \left|F^{2}_{xy}(s,\nu)\right|^{q/2},\nonumber\\
F^{q}_{zz}(s)&=\frac{1}{2N_{s}}\sum_{\nu=1}^{2N_{s}}
  \left[F^{2}_{zz}(s,\nu)\right]^{q/2}.
\label{eq:Fq_bivar}
\end{align}
Following~\cite{Oswiecimka2014,Kwapien2015} these are defined without the $1/q$ root, which is applied only when the scaling is considered; the convention therefore differs from that of Eq.~(\ref{eq:Fq_univar}) for the univariate $F_q(s)$, and we retain it because it is what makes the detrended
cross-correlation coefficient below the exact $q=2$ member of its $q$-dependent generalization. If the two series are cross-correlated in a
scale-invariant manner, the cross-fluctuation function scales as a power law, 
\begin{equation}
\left[F^{q}_{xy}(s)\right]^{1/q}\sim s^{\lambda(q)},
\label{eq:lambda}
\end{equation}
and $\lambda(q)$ is estimated as the slope of a least-squares fit of $\log_{10}F^{q}_{xy}(s)$ against
$\log_{10}s$ over the same scaling range as the univariate analysis, $s\in[18,100]$ weeks for the
Atlantic triplet and the corresponding monthly range for the ONI--TASI pair. For $q=0$ the limit is
evaluated by logarithmic averaging, exactly as in the univariate case. Because the sign-retaining average may change sign or approach zero as $q$ decreases, negative moments are not reliably estimable at this record length; we therefore report $\lambda(q)$ and $\rho(q,s)$ for $q>0$ only, and state this restriction wherever these quantities appear. Uncertainties on $\lambda(q)$ are obtained by the same residual bootstrap of the log--log fit described in Section~\ref{sec_mfdfa} (1000 resamples), and significance is assessed against null bands constructed from 100 pairs of independent IAAFT surrogates. Two diagnostics are applied to $\lambda(q)$, and they answer different questions. The first concerns the multifractality of the coupling itself: following Oświęcimka et al.~\cite{Oswiecimka2014}, a monofractal cross-correlation yields a $\lambda(q)$ that is independent of $q$ and equal to
the exponent obtained from DCCA, whereas a multifractal cross-correlation yields a $\lambda(q)$ that varies with $q$, with the DCCA exponent recovered at $q=2$. We therefore quantify the coupling's multifractality by the range $\Delta\lambda=\max_q\lambda(q)-\min_q\lambda(q)$ and compare it with the
bootstrap uncertainty on $\lambda(q)$ itself. A prior condition is that $F^{q}_{xy}(s)$ actually scales: if it fluctuates around zero rather than
developing a power law, there is no fractal cross-correlation at that $q$~\cite{Oswiecimka2014}, and we verify the sign consistency of
$F^{q}_{xy}(s)$ across the fitted range for every pair and every moment before estimating $\lambda(q)$.

The second diagnostic is the comparison of $\lambda(q)$ with the mean of the individual generalized Hurst exponents, $[h_x(q)+h_y(q)]/2$. This is not a test of cross-multifractality: as~\cite{Oswiecimka2014} shows, the two coincide when the scaling properties of the two series are alike and diverge---in either direction, the sign being fixed by the ratio of the scaling prefactors---as the series become dissimilar. We use it accordingly, as a
measure of how alike the members of each pair are.

At $q=2$ the sign function is inert and Eq.~(\ref{eq:Fq_bivar}) reduces to the mean detrended covariance and the mean detrended variances over all $2N_s$ segments of scale $s$. This is the structure that underlies the detrended cross-correlation coefficient of Zebende~\cite{zebende2011dcca},
\begin{equation}
\rho_{\mathrm{DCCA}}(s)=\frac{F^{2}_{xy}(s)}
{\sqrt{F^{2}_{xx}(s)\,F^{2}_{yy}(s)}},
\label{eq:rhodcca}
\end{equation}
which quantifies the strength of the coupling at the time scale $s$ and is the detrended analogue of the Pearson coefficient. Because every segment enters Eq.~(\ref{eq:rhodcca}) with the same weight, $\rho_{\mathrm{DCCA}}$ cannot distinguish couplings carried by fluctuations of different amplitude. To resolve that, we also use its $q$-dependent generalization~\cite{Kwapien2015}, denoted $\rho_q(s)$ there and written $\rho(q,s)$ here,
\begin{equation}
\rho(q,s)=\frac{F^{q}_{xy}(s)}
{\sqrt{F^{q}_{xx}(s)\,F^{q}_{yy}(s)}},
\label{eq:rhoq}
\end{equation}
in which $q$ acts as a filter on the amplitude of the joint fluctuations: for $q>2$ the segments carrying the largest fluctuations dominate, for $0<q<2$ the smallest ones do, and $q=2$ restores Eq.~(\ref{eq:rhodcca}) exactly.

We report $\lambda(q)$ and $\rho(q,s)$ for $q>0$ only, and state this restriction wherever these quantities appear. The reason is structural rather
than statistical: the Cauchy--Schwarz-type inequality $[F^q_{xy}(s)]^2\le F^q_{xx}(s)F^q_{yy}(s)$, which guarantees $-1\le\rho(q,s)\le1$, holds only for $q\ge0$. For $q<0$ the denominator of Eq.~(\ref{eq:rhoq}) may become arbitrarily small relative to the numerator, so
that $|\rho(q,s)|$ can exceed unity by orders of magnitude and the quantity is no longer interpretable as a correlation coefficient~\cite{Kwapien2015}. The $q\to0$ limit does not arise for the quantities we report; in the univariate case it is evaluated by logarithmic averaging, as described in
Section~\ref{sec_mfdfa}.

Uncertainties on $\lambda(q)$ and $\rho(q,s)$ are obtained by the same residual bootstrap of the log--log fit described in Section~\ref{sec_mfdfa}, and
significance is assessed against null bands constructed from 100 pairs of independent IAAFT surrogates, which preserve the autocorrelation structure of
each series while destroying any mutual coupling. Genuine cross-multifractality requires $\lambda(q)$ to depend on $q$ beyond the component series; the
diagnostic we use throughout is the comparison of $\lambda(q)$ with $[h_{x}(q)+h_{y}(q)]/2$, which coincide when the coupling carries no
multifractality of its own~\cite{Oswiecimka2014}.

For the pairwise coupling among the three Atlantic indices the weekly resolution was retained. For the association with ENSO, both the ONI and the TASI gradient were taken at monthly resolution (the weekly TASI aggregated to monthly means) so as to match the native resolution of the ONI. Rather than fixing a single lag a priori, we computed $\rho_{DCCA}$ at a representative scale as a function of the lag $\tau$ between ONI and TASI, and assessed the significance of the resulting peak with a \emph{family-wise} null: for each pair of independent IAAFT surrogates we recorded the maximum $|\rho_{DCCA}|$ over all lags, so that the significance threshold accounts for the search over lags. This procedure supersedes the directional causal diagnostics (Granger causality, transfer entropy) employed in exploratory analyses, which presuppose a coupling whose statistical basis is not established here (Section~\ref{sec:oni}).

\section{Results and Discussion}
\label{sec:results}

\subsection{Global Multifractal Characterization}
\label{sec:global}

We first characterize the scaling of the three Tropical Atlantic SST indices over the full
observational period. The fluctuation functions $F_q(s)$ follow clean power laws over $s\in[18,100]$
weeks, with $R^{2}\ge0.975$ for every index and every moment of the primary range (minima $0.975$,
$0.991$ and $0.992$ for SAT, TSA and TASI; Supplementary~S1). Beyond $\approx100$ weeks the log--log
slope changes, so that a single power law no longer describes the whole of the accessible range. The
exponents reported here are accordingly estimated over the interval in which scaling actually holds;
what happens when the fit is extended past that point is examined in Section~\ref{sec:origin}, and it
is not a second-order correction.

\begin{figure}[htbp]
    \centering
    \begin{subfigure}[b]{0.48\textwidth}
        \centering
        \includegraphics[width=\textwidth]{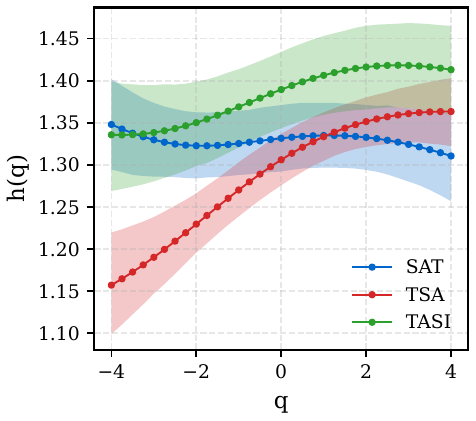}
        \caption{}
        \label{fig:hq}
    \end{subfigure}
    \hfill
    \begin{subfigure}[b]{0.48\textwidth}
        \centering
        \includegraphics[width=\textwidth]{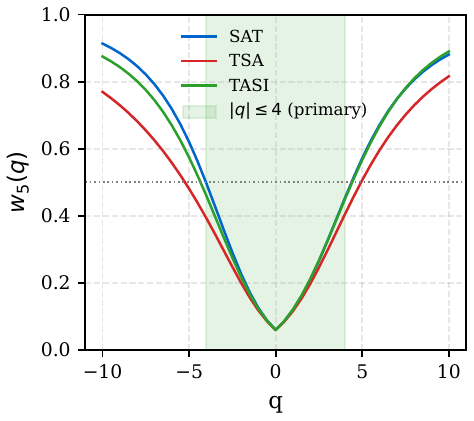}
        \caption{}
        \label{fig:conc}
    \end{subfigure}

    \caption{Apparent multifractal characterization of the Tropical Atlantic SST indices produced by
    the conventional protocol (second-order detrending, $s\in[18,100]$ weeks).
    (a)~Generalized Hurst exponent $h(q)$ over the primary moment range $|q|\le4$, with $95\%$
    bootstrap bands. The apparent widths are $\Delta h=0.217$ (TSA), $0.077$ (SAT) and $0.054$
    (TASI), and the bands overlap over most of the range. None of the three curves is monotonically
    decreasing: $h(q)$ increases monotonically for TSA, increases up to $q\approx0$ and decreases
    thereafter for TASI, and is non-monotonic in the interior for SAT. Since
    $f(\alpha)=1+q^{2}h'(q)$, every interval on which $h(q)$ increases yields $f(\alpha)>1$, which no
    one-dimensional record can attain; the singularity spectrum is therefore not reported as a
    result.
    (b)~Concentration of the moments: $w_5(q)$ is the fraction of the fluctuation function $F_q(s)$
    carried by the five largest segments at each scale, averaged over the fitted range; the shaded
    band marks the range $|q|\le4$ used in the primary analysis. Beyond $|q|\approx5$ the moments are
    increasingly dominated by a handful of extreme segments, so that $h(q)$ there reflects the upper
    tail of the segment-variance distribution rather than a scaling property of the record.}
    \label{fig:h_tau_f}
\end{figure}

Figure~\ref{fig:h_tau_f}(a) shows the generalized Hurst exponents obtained from this range over the
primary moment range $|q|\le4$. The apparent widths are $\Delta h=0.217$ for TSA, $0.077$ for SAT and
$0.054$ for TASI. Three features of this figure qualify those numbers before any of them is
interpreted.

First, they are not well determined at this record length. The bootstrap bands of the three indices
overlap over most of the moment range, and the confidence interval of every width includes values
several times smaller than the point estimate (Table~\ref{tab:surrogates}). Second, the moment range
in most widespread use is not supported by the data. Figure~\ref{fig:h_tau_f}(b) shows the fraction of
$F_q(s)$ carried by the five largest segments: beyond $|q|\approx5$ the moments are progressively
dominated by a handful of extreme segments, so that $h(q)$ in that regime measures the upper tail of
the segment-variance distribution rather than a scaling property of the record. Extending the range to
$|q|\le10$ accordingly inflates the widths---by factors of $3.8$, $1.0$ and $2.5$ for SAT, TSA and
TASI---and reorders the three indices while doing so.

The third feature is qualitative and has the sharpest consequence. None of the three curves is
monotonically decreasing in $q$, as a multifractal spectrum requires. The exponent of TSA increases
monotonically over the whole primary range, from $h(-4)\approx1.16$ to $h(+4)\approx1.38$; that of
TASI increases up to $q\approx0$ and decreases thereafter; that of SAT decreases, rises over an
interior interval, and decreases again. Combining the two relations of Section~\ref{sec_mfdfa},
$\alpha=\mathrm{d}\tau/\mathrm{d}q=h+qh'$ and $f(\alpha)=q\alpha-\tau$ with $\tau=qh-1$, gives
\begin{equation}
f(\alpha)=1+q^{2}h'(q),
\label{eq:falpha_identity}
\end{equation}
so that $f(\alpha)\le1$ if and only if $h'(q)\le0$. Every interval on which $h(q)$ increases therefore
yields a singularity spectrum exceeding unity---a fractal dimension larger than the dimension of the
support, which a one-dimensional record cannot attain. All three indices violate the bound on some
subinterval of $|q|\le4$: severely for TSA, where the tilt is systematic and $f$ reaches
$\approx1.13$; over the negative moments for TASI; and marginally, over a short interior interval,
for SAT. At the wider moment range the transform fails in a second and independent way, $\alpha(q)$
ceasing to be monotonic at between $7$ and $12$ of $40$ grid points depending on the index, so that
$\tau(q)$ is not concave and the transform folds back on itself.

We therefore report no singularity spectrum. This is not a presentational choice: Eq.~(\ref{eq:falpha_identity}) shows that the spectra these records
yield are not admissible objects, and smoothing or truncating them would conceal the diagnostic
rather than repair it. What $f(\alpha)$ retains is exactly that diagnostic value, as a consistency
check on $h(q)$: a spectrum that exceeds unity is a direct indication that the estimated $h(q)$ is not
that of a multifractal, whether because the width is finite-size scatter or because the fit has been
performed across a change of scaling regime. It is worth being precise about what this does and does
not establish. A degenerate spectrum is compatible with a monofractal record whose $h(q)$ carries
finite-size scatter, but it does not by itself demonstrate monofractality: a genuine multifractal of
this length would produce a comparably ill-behaved transform. The classification of these records is
settled in Section~\ref{sec:origin} by comparison against surrogate ensembles, not here.

The apparent widths of Figure~\ref{fig:h_tau_f}(a) accordingly serve only to motivate that comparison.
None of them is taken as evidence of multifractality, and the ordering they suggest is not interpreted
at all---as Section~\ref{sec:origin} shows, it does not survive the choice of fitting range, moment
range or detrending order.

\subsection{Origin of the multifractality: surrogate analysis}
\label{sec:origin}

To decide whether the apparent widths of Section~\ref{sec:global} reflect genuine nonlinear dynamics
or are an artefact of the amplitude distribution and of the linear (two-point) correlations, we compare
each record against two surrogate ensembles~\cite{schreiber1996improved,Kantelhardt_2002}. Random
shuffling destroys all temporal correlations while preserving the amplitude distribution, and therefore
measures the finite-size floor against which any genuine width must be judged. The iterative
amplitude-adjusted Fourier transform (IAAFT) additionally preserves the linear autocorrelation, i.e.\
the power spectrum, so that any width in excess of the IAAFT ensemble can only originate in nonlinear
temporal correlations~\cite{KwapienDrozdz2012}. For each index we generated $N_{s}=100$ realizations of
each surrogate type, estimated $h(q)$ and the width $\Delta h=\max_{q}h(q)-\min_{q}h(q)$ for every
realization, and obtained the ensemble mean, its standard deviation, and a one-sided empirical
$p$-value. Confidence intervals for the original series were obtained by a residual bootstrap of the
log--log fit that resamples the same scales across all moments, thereby preserving the covariance
between moments.

Two methodological choices govern what follows, and both are fixed before any width is interpreted.
The first is the fitting range. We estimate all exponents over $s\in[18,100]$ weeks. The lower bound is
the smallest scale at which a second-order detrending is well determined, $s_{\min}\gtrsim6(m+1)=18$
for $m=2$; below it the polynomial overfits the individual segments, and the resulting fluctuation
functions are not estimates of anything. The upper bound stops short of the change of slope that
$F_q(s)$ displays beyond $\approx100$ weeks (Supplementary~S1). This matters more than it may appear.
A single power law fitted across a change of scaling regime does not measure the $q$-dependence of the
exponent \emph{within} a regime: it measures how the exponents of two regimes combine, and that
combination is itself $q$-dependent, so the fit returns a spurious width whose sign and magnitude
depend on the record. The second choice is the moment range, restricted to $|q|\le4$. At this record
length $F_q(s)$ at large $|q|$ is governed by a handful of extreme segments
(Fig.~\ref{fig:h_tau_f}(b)), so that $h(q)$ for $|q|\gtrsim5$ measures the upper tail of the
segment-variance distribution rather than a scaling property~\cite{Drozdz2009,Zhou2012}.

\begin{figure}[t]
   \centering
   \includegraphics[width=0.32\textwidth]{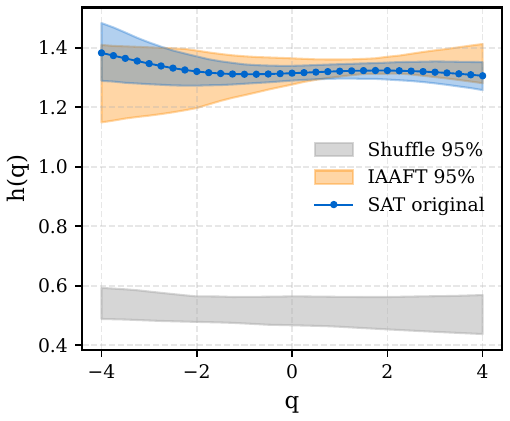}\hfill
   \includegraphics[width=0.32\textwidth]{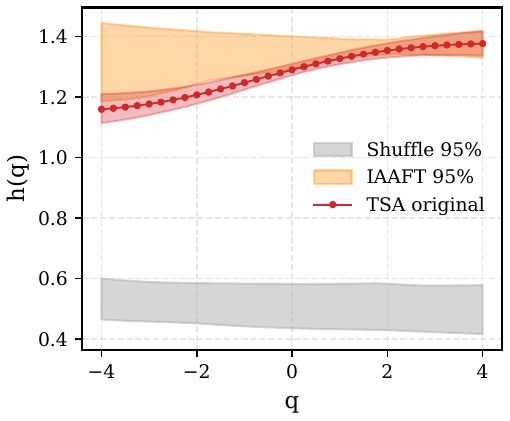}\hfill
   \includegraphics[width=0.32\textwidth]{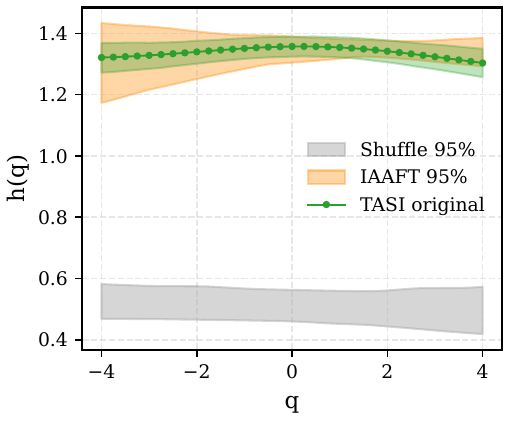}
   \caption{Generalized Hurst exponents $h(q)$ of the original series (markers, with $95\%$ bootstrap
   band) against the $95\%$ envelopes of the shuffled (grey) and IAAFT (orange) surrogate ensembles;
   DFA-2, $s\in[18,100]$ weeks, $|q|\le4$, $N_{s}=100$ realizations of each ensemble. SAT and TASI lie
   within their IAAFT envelopes at every moment, consistent with the $p$-values of
   Table~\ref{tab:surrogates}. TSA departs from its envelope, but at the negative moments and in the
   direction of a \emph{smaller} exponent, running at or below the lower edge of the envelope for
   $q<0$ and merging with it for $q>0$---the opposite of the pattern produced by a multiplicative
   cascade, whose excess lies above the envelope at $q<0$. The width
   $\Delta h=\max_{q}h(q)-\min_{q}h(q)$ is blind to this distinction, since as a range it discards the
   sign of $h'(q)$. The shuffled envelopes lie near $h\approx0.5$ for all three indices, well below the
   IAAFT envelopes near $h\approx1.3$, as they must: the IAAFT ensemble preserves the linear
   autocorrelation that shuffling destroys.}
   \label{fig:hq_bands}
\end{figure}

Table~\ref{tab:surrogates} reports the outcome, and Figure~\ref{fig:hq_bands} shows it directly.
Within the primary range the widths of SAT and TASI ($\Delta h=0.077$ and $0.054$) lie inside both
surrogate ensembles---indeed below the IAAFT means of $0.097$ and $0.070$---and are not significant
against either null; their $h(q)$ curves fall within the IAAFT envelope at every moment. Their
apparent multifractality is fully accounted for by their linear correlation structure and the finite
sample size. The width of TSA, $\Delta h=0.217$, is the largest of the three and exceeds both the
shuffled ensemble ($p=0.010$) and, nominally, the IAAFT ensemble ($p=0.040$).

That excess, however, does not survive an internal consistency check, and the check is structural
rather than statistical. The generalized Hurst exponent of TSA \emph{increases} monotonically with $q$
over the primary range, from $h(-4)\approx1.16$ to $h(+4)\approx1.38$ (Fig.~\ref{fig:h_tau_f}(a)).
Combining the two relations of Section~\ref{sec_mfdfa}, $\alpha=\mathrm{d}\tau/\mathrm{d}q=h+qh'$ and
$f(\alpha)=q\alpha-\tau$, gives the identity
\begin{equation}
f(\alpha)=1+q^{2}h'(q),
\label{eq:falpha_identity_1}
\end{equation}
so that $f(\alpha)\le1$ if and only if $h'(q)\le0$. An increasing $h(q)$ therefore forces
$f(\alpha)>1$ over the range in which it increases---a fractal dimension exceeding the dimension of
the support, which no one-dimensional record can attain. This is not a numerical artefact of
differentiating a noisy $\tau(q)$: it is an algebraic consequence of the sign of $h'$, and it is what
produces the value $f\approx1.13$ reached by TSA. The width of TSA is thus not a weak or marginal
multifractal signature but a quantity of the wrong sign: a multiplicative cascade generates
$h'(q)<0$, whereas what is measured here is the opposite. Figure~\ref{fig:hq_bands} shows this directly: the TSA curve departs from its IAAFT envelope at the negative moments, where it runs at or below the lower edge, and merges with the envelope at the
positive ones. A multiplicative cascade produces the reverse---an excess above the envelope at $q<0$,
where the small fluctuations dominate. The statistic itself does not distinguish the two cases: as a
range, $\Delta h$ discards the sign of $h'$, so a curve tilted the wrong way enters it exactly as a
genuine cascade would. This is a further reason to treat an apparent width as a summary to be checked
against the shape of $h(q)$ rather than as a result in itself.Two further observations point the same
way. The width appears only once the detrending order reaches two---$\Delta h_{\mathrm{TSA}}$ rises
from $0.040$ under DFA-1 to $0.217$ under DFA-2 (Table~\ref{tab:detrending})---and a cascade is not
created by removing a parabola from each segment. And it is the only width in the analysis that is
insensitive to the moment range (a factor of $1.0$ between $|q|\le4$ and $|q|\le10$), which is what a
systematic tilt of $h(q)$ produces, as distinct from the growth with $|q|$ characteristic of
finite-size scatter. We therefore do not interpret the apparent width of TSA as multifractality, and
note that the conclusion does not depend on which $p$-value is adopted as primary: $p=0.040$ and
$p^{+}=0.059$ straddle the conventional threshold in opposite directions, and the argument rests on
the sign of $h'$ rather than on either.

The second block of Table~\ref{tab:surrogates} shows what is at stake in the choice of fitting range.
Extended to $s\in[18,230]$ weeks---across the change of slope---the width of TASI rises from $0.054$
to $0.239$ and becomes significant against both nulls ($p=0.010$, $p^{+}=0.010$), while that of TSA
collapses from $0.217$ to $0.010$ and becomes the least significant quantity in the analysis
($p=1.000$). Neither number describes a property of the corresponding record. Both are the signature
anticipated above: a single exponent fitted across two regimes mixes their slopes in a $q$-dependent
way, and the sign of the resulting bias depends on how the two regimes are arranged in each index. In
the range over which the power law actually holds, the width of TASI is $0.054$ and is compatible with
both nulls.

More generally, the apparent widths are not identifiable at this record length, and this is the
central quantitative result of the analysis. With $N\approx2300$ and about three-quarters of a decade
of usable scales, the ordering of the three indices by $\Delta h$ depends on every discretionary
choice in the protocol. It reverses with the fitting range, as just shown. It reverses with the
moment range: under DFA-2 the widest index is TSA at $|q|\le4$ and SAT at $|q|\le10$
($0.292$ against $0.224$ and $0.137$). And it reverses with the detrending order: the widest index is
TASI under DFA-1 ($0.236$) and TSA under DFA-2 and DFA-3 (Table~\ref{tab:detrending}). Each of the
three indices is accordingly the most multifractal of the three under some combination of choices
that is defensible on its own terms---every $s_{\min}$ used here satisfies the condition imposed by
its detrending order, both moment ranges are in current use, and all three detrending orders are
standard. Within each configuration the bootstrap confidence intervals of the three indices overlap.
No ranking survives, and none should be reported.

Two properties of the records are, by contrast, robust to all of these choices. The first is the
persistence: $h(2)$ lies between $1.19$ and $1.42$ and exceeds unity for every index at every
detrending order by at least seven bootstrap standard errors (Table~\ref{tab:detrending}), so the
series are strongly persistent and locally non-stationary on the analysed scales. This is what
motivates the second-order detrending adopted as primary: a first-order detrending does not fully
remove the low-frequency trends, which leak into $F_q(s)$ in a $q$-dependent way. The second is the
shape of the increment distribution: the weekly increments are mildly leptokurtic, with excess
kurtosis of $0.87$, $0.55$ and $0.61$ for SAT, TSA and TASI (Supplementary~S3), placing them far from
the heavy-tailed regime in which the distribution of fluctuations appreciably broadens a singularity
spectrum~\cite{Kwapien2023,Rak2025}. This bears on the moment range: with increments this close to
Gaussian, the dominance of a few segments at $|q|\gtrsim5$ reflects the finite number of segments
available at the largest scales rather than a genuinely heavy-tailed generating process, so the
widths measured there are finite-size scatter.

In summary, none of the three indices exhibits multifractality that survives comparison with
surrogate ensembles within the resolution afforded by the record. The two regional indices and the
interhemispheric gradient are monofractal on the scales over which a power law holds; the one
nominally significant width, that of TSA, is incompatible with a singularity spectrum bounded by
unity and is attributable to a fit performed across a change of scaling regime. The dynamical
distinction between regional variability and the interhemispheric gradient, which earlier estimates
located in the multifractal width, is not supported at this record length.

\begin{table}[t]
\centering
\small
\setlength{\tabcolsep}{5pt}
\caption{Apparent multifractal width $\Delta h$ of the original series compared with shuffled and
IAAFT surrogate ensembles ($N_{s}=100$ realizations each), DFA-2 and $|q|\le4$, for the primary
fitting range and for the wider range that extends past the change of slope of the fluctuation
functions. Surrogate columns give the ensemble mean $\pm$ one standard deviation; $p$ is the
one-sided empirical probability $P(\Delta h_{\mathrm{surr}}\ge\Delta h_{\mathrm{orig}})$, and $p^{+}$
the same probability for the width restricted to non-negative moments,
$\Delta h^{+}=\max_{q\ge0}h(q)-\min_{q\ge0}h(q)$. Within the primary range only TSA exceeds either
null, and it is excluded on structural grounds in the text. Extending the range past the change of
slope reverses the picture entirely: the width of TASI triples and becomes significant against both
nulls, while that of TSA collapses by a factor of twenty.}
\label{tab:surrogates}
\begin{tabular}{lcccccc}
\toprule
Index & $\Delta h$ (original) & $\Delta h$ shuffled & $p_{\mathrm{shuf}}$
      & $\Delta h$ IAAFT & $p_{\mathrm{IAAFT}}$ & $p^{+}_{\mathrm{IAAFT}}$ \\
      & [95\% CI] & & & & & \\
\midrule
\multicolumn{7}{l}{\emph{Primary range}, $s\in[18,100]$ weeks} \\
SAT   & $0.077\ [0.026,\,0.181]$ & $0.049\pm0.030$ & $0.198$
      & $0.097\pm0.058$ & $0.564$ & $0.792$ \\
TSA   & $0.217\ [0.159,\,0.274]$ & $0.051\pm0.025$ & $0.010$
      & $0.095\pm0.057$ & $0.040$ & $0.059$ \\
TASI  & $0.054\ [0.035,\,0.092]$ & $0.043\pm0.028$ & $0.337$
      & $0.070\pm0.044$ & $0.604$ & $0.129$ \\
\midrule
\multicolumn{7}{l}{\emph{Extended range}, $s\in[18,230]$ weeks} \\
SAT   & $0.111\ [0.039,\,0.182]$ & $0.036\pm0.022$ & $0.010$
      & $0.095\pm0.052$ & $0.337$ & $0.624$ \\
TSA   & $0.010\ [0.009,\,0.109]$ & $0.037\pm0.023$ & $0.901$
      & $0.087\pm0.046$ & $1.000$ & $0.980$ \\
TASI  & $0.239\ [0.161,\,0.321]$ & $0.036\pm0.022$ & $0.010$
      & $0.105\pm0.053$ & $0.010$ & $0.010$ \\
\bottomrule
\end{tabular}
\end{table}

\begin{table}[htbp]
\centering
\small
\setlength{\tabcolsep}{5pt}
\begin{tabular}{lccccc}
\toprule
Index & $h(2)$ & $\Delta h$ ($|q|\le4$) & $\Delta h$ ($|q|\le10$) & ratio & $R^{2}_{\min}$ \\
\midrule
\multicolumn{6}{l}{\emph{DFA-1}, $s\in[12,100]$ weeks} \\
SAT  & $1.192\pm0.026$ & $0.099\ [0.037,\,0.166]$ & $0.260$ & $2.6$ & $0.988$ \\
TSA  & $1.228\pm0.023$ & $0.040\ [0.014,\,0.111]$ & $0.077$ & $1.9$ & $0.988$ \\
TASI & $1.192\pm0.027$ & $\mathbf{0.236}\ [0.143,\,0.350]$ & $0.400$ & $1.7$ & $0.982$ \\
\midrule
\multicolumn{6}{l}{\emph{DFA-2}, $s\in[18,100]$ weeks (primary analysis)} \\
SAT  & $1.323\pm0.016$ & $0.077\ [0.026,\,0.181]$ & $0.292$ & $3.8$ & $0.975$ \\
TSA  & $1.354\pm0.013$ & $\mathbf{0.217}\ [0.159,\,0.274]$ & $0.224$ & $1.0$ & $0.991$ \\
TASI & $1.342\pm0.019$ & $0.054\ [0.035,\,0.092]$ & $0.137$ & $2.5$ & $0.992$ \\
\midrule
\multicolumn{6}{l}{\emph{DFA-3}, $s\in[24,100]$ weeks} \\
SAT  & $1.333\pm0.021$ & $0.037\ [0.021,\,0.102]$ & $0.264$ & $7.1$ & $0.992$ \\
TSA  & $1.351\pm0.016$ & $\mathbf{0.206}\ [0.148,\,0.269]$ & $0.283$ & $1.4$ & $0.987$ \\
TASI & $1.416\pm0.026$ & $0.083\ [0.027,\,0.158]$ & $0.082$ & $1.0$ & $0.989$ \\
\bottomrule
\end{tabular}
\caption{Dependence of the generalized Hurst exponent $h(2)$ and of the apparent multifractal width
$\Delta h$ on the detrending order. The minimum scale is set consistently with the detrending order
($s_{\min}\gtrsim6(m+1)$ for order $m$) and $s_{\max}=100$ weeks throughout, so that the three orders
are compared on scales at which each is well determined; $h(2)$ carries its bootstrap standard error
and $\Delta h$ its $95\%$ bootstrap confidence interval. The ratio column is
$\Delta h(|q|\le10)/\Delta h(|q|\le4)$. The persistence is order-independent in the only sense the
analysis requires: $h(2)$ exceeds unity for every index at every order by at least seven standard
errors, so the diagnosis of local non-stationarity---and with it the inadequacy of a first-order
detrending---does not depend on the order adopted. Its magnitude, however, does: $h(2)$ rises by
between $0.12$ and $0.22$ from DFA-1 to DFA-3, five to ten times the bootstrap standard errors, and
this spread persists even though $s_{\min}$ has been raised in step with the order. The apparent width
is not order-independent in any sense: the widest index (bold) is TASI under DFA-1 and TSA under DFA-2
and DFA-3, the confidence intervals overlap across indices within every order, and widening the moment
range gives a third ordering again. No ranking of the three indices by apparent multifractal strength
survives these choices.}
\label{tab:detrending}
\end{table}

\subsection{Time-resolved analysis: local Hurst exponent and stationarity of the multifractality}
\label{sec:windows}

We analysed the record in moving windows (Section~\ref{sec:moving_window}) to test whether the scaling properties vary in time. Because a five-year window does not support a reliable estimate of the full multifractal spectrum, we take the local Hurst exponent $h_2(t)$ as the primary time-resolved measure and treat the local width $\Delta h(t)$ only in conjunction with a window-specific surrogate test.

\begin{figure}[htbp]
    \centering
    \includegraphics[width=0.85\linewidth]{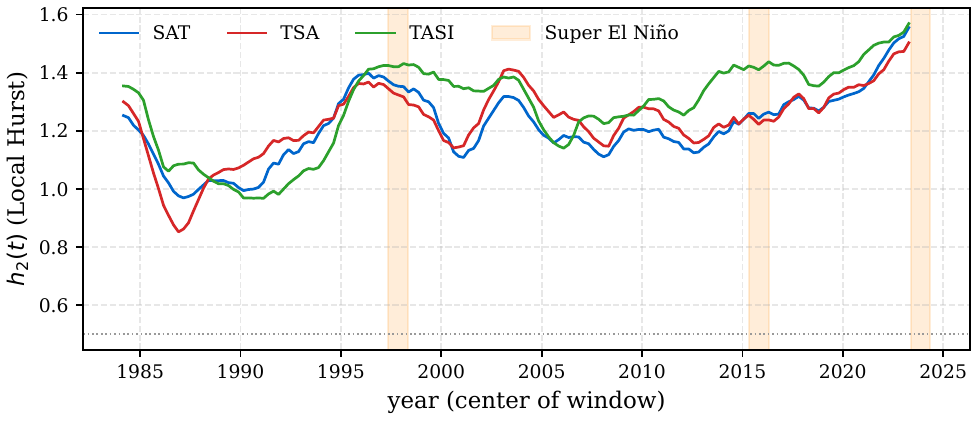}
    \caption{Local Hurst exponent $h_2(t)$ (five-year windows, DFA-2) for the three indices; shaded bands mark the strong El Ni\~no events. All indices remain persistent ($h_2>1$) throughout, with jointly varying strength.}
    \label{fig:h2t}
\end{figure}

\begin{figure}[htbp]
    \centering
    \begin{subfigure}[b]{0.32\textwidth}\centering\includegraphics[width=\textwidth]{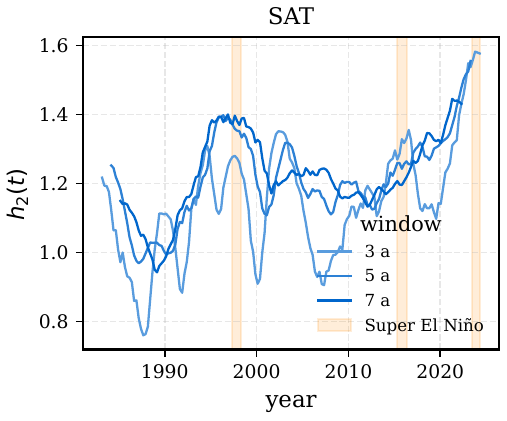}\caption{}\end{subfigure}\hfill
    \begin{subfigure}[b]{0.32\textwidth}\centering\includegraphics[width=\textwidth]{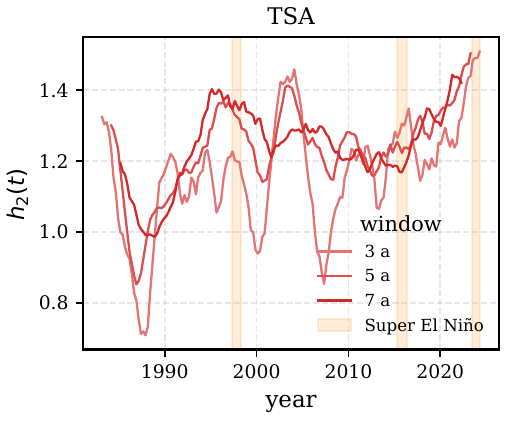}\caption{}\end{subfigure}\hfill
    \begin{subfigure}[b]{0.32\textwidth}\centering\includegraphics[width=\textwidth]{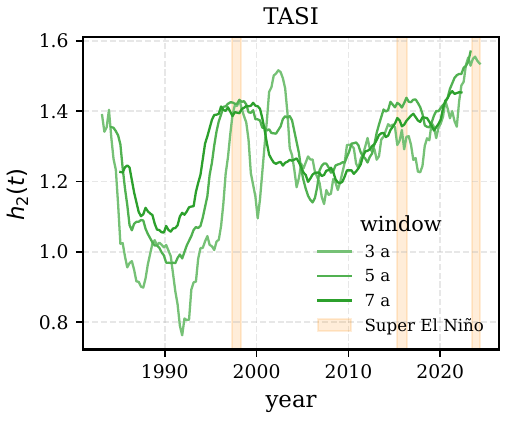}\caption{}\end{subfigure}
    \caption{Robustness of $h_2(t)$ to the window length (three, five and seven years) for SAT, TSA and TASI. The trajectories share the same low-frequency structure.}
    \label{fig:h2t_sens}
\end{figure}

The local Hurst exponent (Fig.~\ref{fig:h2t}) confirms that all three indices remain strongly persistent throughout the record, with $h_2(t)$ fluctuating between $\approx 1.0$ and $\approx 1.5$ (Table~\ref{tab:windows}); the persistence weakens around 1987--1991 and strengthens after $\sim 2010$ in the three series jointly. These variations are robust to the window length: the $h_2(t)$ trajectories obtained with three-, five- and seven-year windows share the same low-frequency structure, the shorter windows being noisier (Fig.~\ref{fig:h2t_sens}).

\begin{figure}[htbp]
    \centering
    \begin{subfigure}[b]{0.32\textwidth}\centering\includegraphics[width=\textwidth]{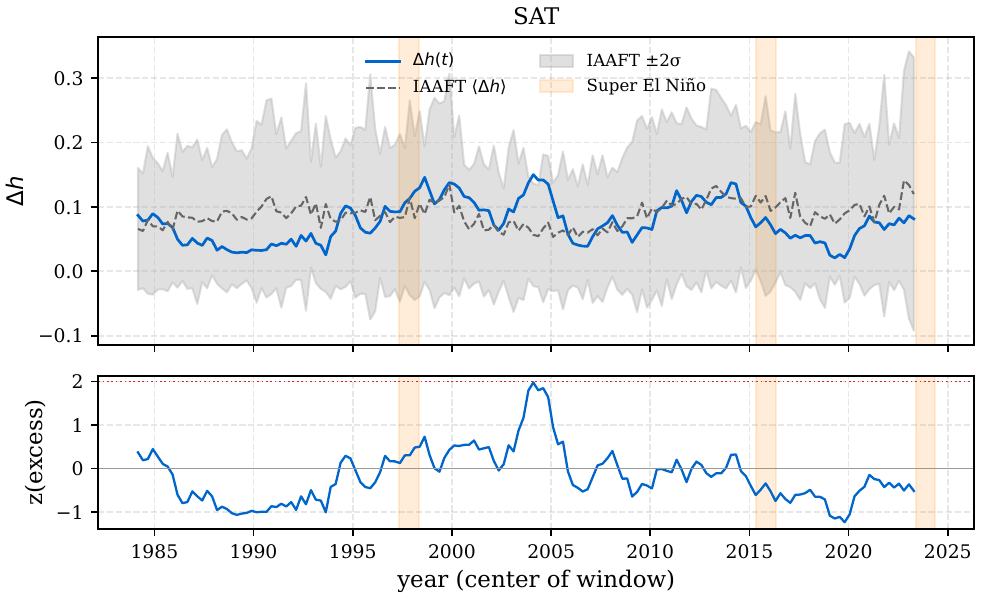}\caption{}\end{subfigure}\hfill
    \begin{subfigure}[b]{0.32\textwidth}\centering\includegraphics[width=\textwidth]{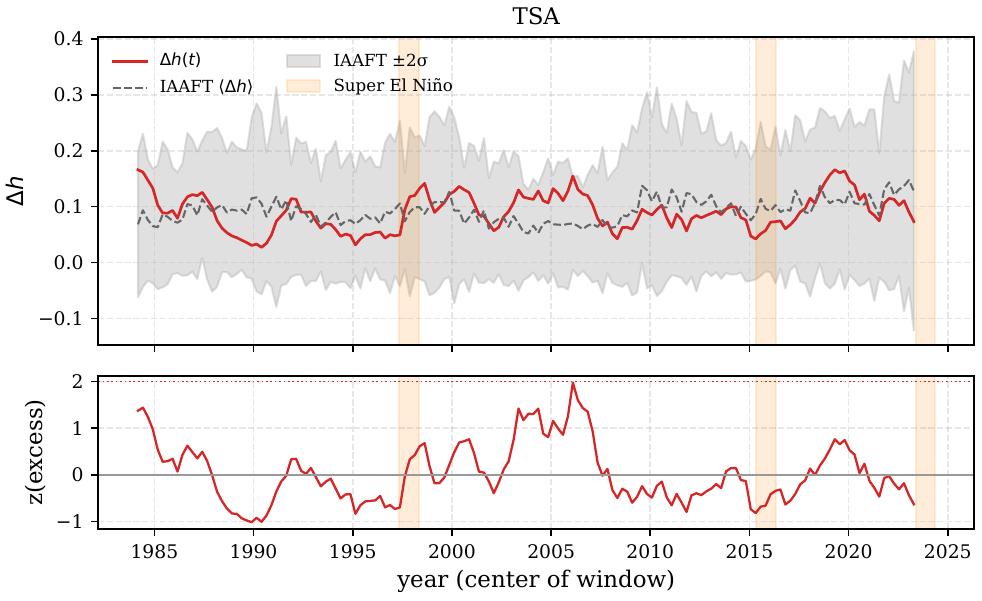}\caption{}\end{subfigure}\hfill
    \begin{subfigure}[b]{0.32\textwidth}\centering\includegraphics[width=\textwidth]{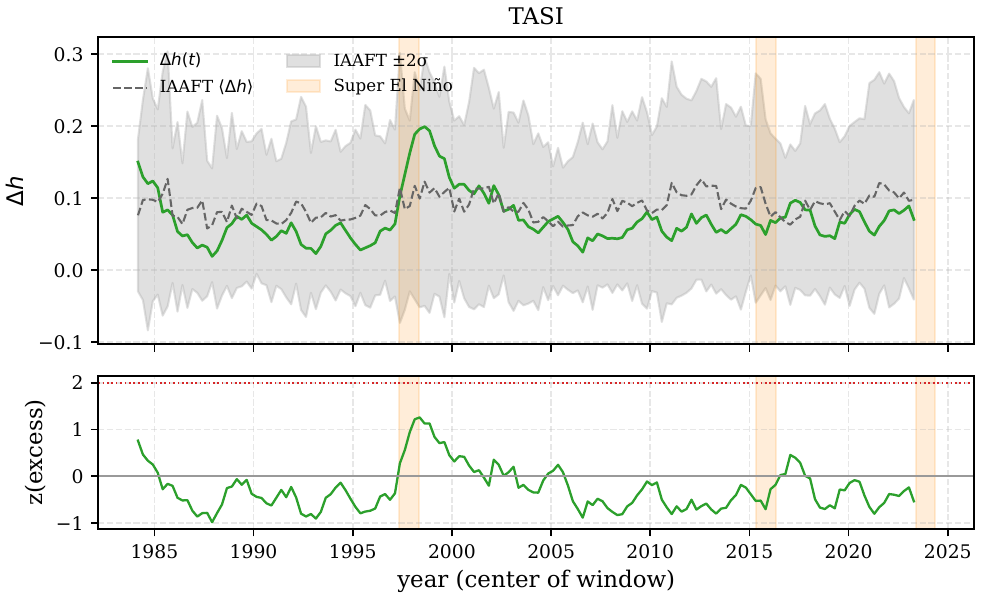}\caption{}\end{subfigure}
    \caption{Local multifractal width $\Delta h(t)$ (top of each panel) against the mean and $\pm 2\sigma$ band of $K=50$ local IAAFT surrogates, with the standardized excess $z(t)$ below (dotted line at $z=2$). No index sustains $z>2$; the excess is at chance level for all three.}
    \label{fig:dh_excess}
\end{figure}

In contrast, the multifractal width shows no evidence of time localization. The fraction of windows in which $\Delta h(t)$ exceeds its local IAAFT ensemble at the $5\%$ level is $1.9\%$, $5.7\%$ and $1.3\%$ for SAT, TSA and TASI respectively---at or below the nominal false-positive rate (Fig.~\ref{fig:dh_excess}). Because consecutive windows overlap by $95\%$, the record contains only $\sim 8$ independent windows, so these fractions are fully consistent with the absence of any localized effect; the standardized excess remains modest for every index (95th percentile of the $z$-score below $1.9$, and lowest for TASI). The same picture holds for all three window lengths. The two isolated windows in which TASI is nominally significant bracket the 1997--1998 event, but they are too few and too isolated to establish a relationship between the gradient's multifractality and El Ni\~no.

These results are consistent with the global analysis of Section~\ref{sec:origin} once statistical power is taken into account. The genuine multifractality of TASI corresponds to a small excess over the surrogate level ($\Delta h - \langle\Delta h\rangle_{\mathrm{IAAFT}} \approx 0.10$ over the full record) that resides in the long-range correlations up to scales of several years; within a five-year window this excess is smaller than the surrogate scatter ($\pm 2\sigma \approx 0.1$) and cannot be resolved. The multifractality of TASI is therefore a stationary, record-spanning property rather than an episodic phenomenon, and none of the three indices exhibits statistically significant time-varying or event-driven multifractality. The time dependence that is robustly established is that of the persistence $h_2(t)$, not of the multifractal width.

\begin{table}[t]
\centering
\caption{Moving-window analysis (five-year windows, step $13$ weeks, second-order detrending). $h_2(t)$ is summarized by its median and central $90\%$ range; the last column gives the fraction of windows in which $\Delta h(t)$ exceeds the local IAAFT ensemble at the $5\%$ level. With $\sim 8$ independent windows these fractions are consistent with chance for all indices.}
\label{tab:windows}
\begin{tabular}{lccc}
\toprule
Index & median $h_2(t)$ & $h_2(t)$ [5--95\%] & windows with $p<0.05$ \\
\midrule
SAT  & $1.21$ & $[1.00,\,1.42]$ & $1.9\%$ \\
TSA  & $1.23$ & $[1.00,\,1.43]$ & $5.7\%$ \\
TASI & $1.33$ & $[0.99,\,1.47]$ & $1.3\%$ \\
\bottomrule
\end{tabular}
\end{table}

\subsection{Cross-correlations between the indices}
\label{sec:cross}

The pairwise coupling of the indices was quantified with the coefficients defined in Section~\ref{sec:entropy_and_mfdcca}: the detrended cross-correlation coefficient $\rho_{\mathrm{DCCA}}(s)$ of Eq.~(\ref{eq:rhodcca}), its $q$-dependent generalization $\rho(q,s)$ of Eq.~(\ref{eq:rhoq}), and the MFCCA cross-scaling exponent $\lambda(q)$ of Eq.~(\ref{eq:lambda}). As stated there, $\rho(q,s)$ and $\lambda(q)$ are reported for $q>0$.

\begin{figure}[htbp]
    \centering
    \begin{subfigure}[b]{0.48\textwidth}
        \centering
        \includegraphics[width=\textwidth]{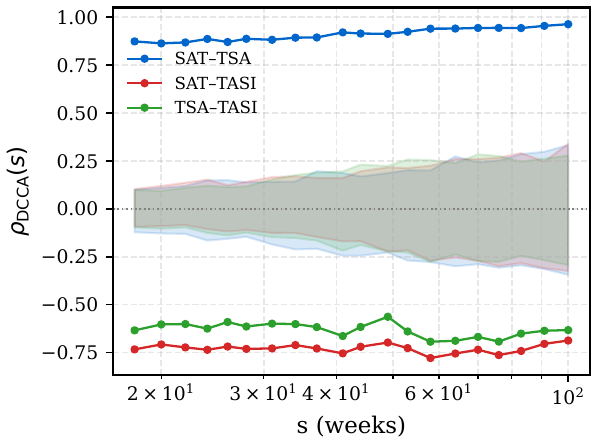}
        \caption{}
        \label{fig:rho_s}
    \end{subfigure}
    \hfill
    \begin{subfigure}[b]{0.48\textwidth}
        \centering
        \includegraphics[width=\textwidth]{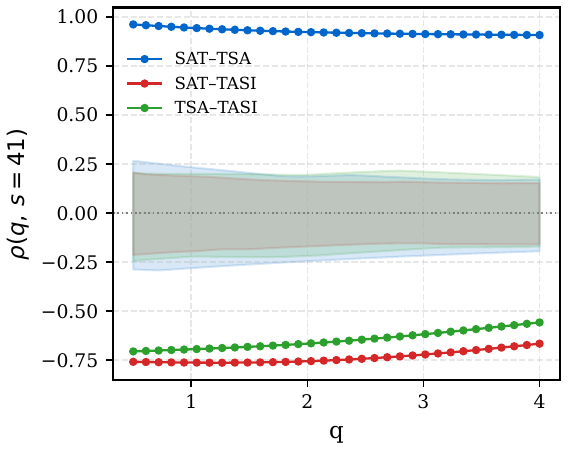}
        \caption{}
        \label{fig:rho_q}
    \end{subfigure}
    \caption{Detrended cross-correlation between the three indices, cut two ways. (a)~$\rho_{\mathrm{DCCA}}(s)$, Eq.~(\ref{eq:rhodcca}), against the
    time scale. (b)~The $q$-dependent coefficient $\rho(q,s)$, Eq.~(\ref{eq:rhoq}), against the moment order at the fixed scale $s=52$
    weeks, the geometric mean of the fitted range. The two panels are orthogonal sections through the same quantity: panel (a) is the $q=2$
    section of $\rho(q,s)$ and panel (b) its $q$-dependence at one scale. Shaded bands are the $95\%$ null envelopes from 100 pairs of independent
    IAAFT surrogates, which preserve the autocorrelation of each series while destroying their mutual coupling; they widen with $s$ in panel (a) as the
    number of available segments falls. In panel (b), $q$ acts as a filter on the amplitude of the joint fluctuations---$q>2$ weights the segments
    carrying the largest fluctuations, $0<q<2$ the smallest---so that the mild downward slope of the TASI pairs indicates a coupling slightly weaker among
    the largest joint excursions than among the smallest. The dependence is smooth and monotonic in all three pairs and does not by itself indicate
    multifractal cross-correlation, which is diagnosed instead from the constancy of $\lambda(q)$ (Fig.~\ref{fig:lambda}).}
    \label{fig:rhodcca}
\end{figure}

The two regional indices SAT and TSA are strongly and positively cross-correlated at all scales, the
coefficient rising from $\rho_{\mathrm{DCCA}}=+0.89$ at $s=24$~weeks to $+0.94$ at $s=76$~weeks and
$+0.96$ at the longest fitted scale (Fig.~\ref{fig:rhodcca}, Table~\ref{tab:cross}). In contrast, the interhemispheric
gradient TASI is strongly anti-correlated with both regional indices, with a coefficient that is
essentially scale-independent over the whole range: $\rho_{\mathrm{DCCA}}$ lies between $-0.78$ and
$-0.69$ for SAT--TASI and between $-0.69$ and $-0.56$ for TSA--TASI. Every scale is individually significant against the IAAFT null for all three pairs, at the floor $p=0.010$ attainable with $100$ surrogate pairs: the observed coefficients lie far outside a null band
that spans roughly $\pm0.15$ at the shortest scales and widens to $\pm0.30$ at the longest. This pattern is physically coherent: the two regional
sea-surface-temperature anomalies co-vary, and increasingly so towards longer time scales, whereas the
gradient index---by construction a contrast between hemispheres---opposes them.

\begin{figure}[t]
   \centering
   \includegraphics[width=0.32\textwidth]{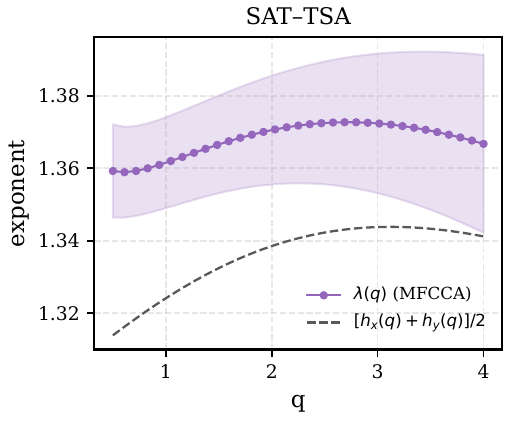}\hfill
   \includegraphics[width=0.32\textwidth]{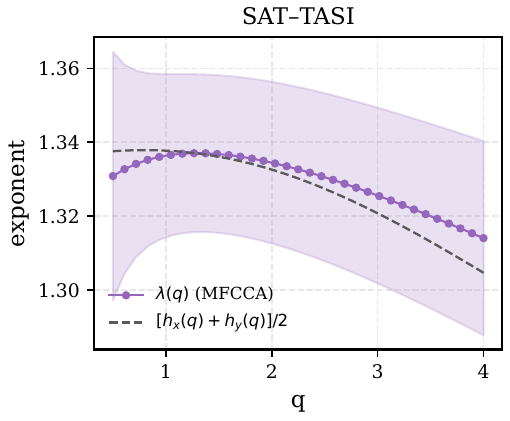}\hfill
   \includegraphics[width=0.32\textwidth]{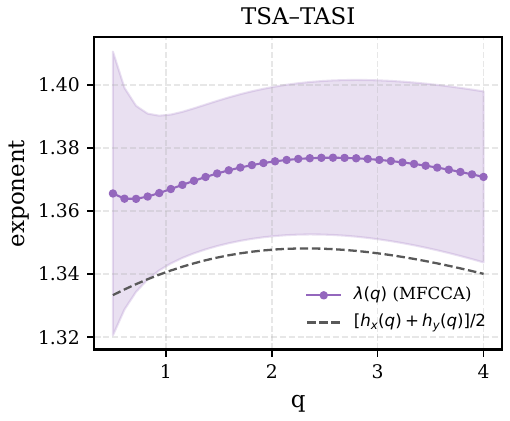}
   \caption{MFCCA cross-scaling exponents $\lambda(q)$ (markers, $q>0$) with bootstrap uncertainties, compared with $[h_x(q)+h_y(q)]/2$ (dashed) for each pair. Monofractality of the coupling is indicated by the constancy of $\lambda(q)$ in $q$~\cite{Oswiecimka2014}, which holds for all three pairs to within the uncertainty on a single exponent ($\Delta\lambda=0.027$, $0.014$ and $0.035$ against standard errors of $0.02$--$0.04$). The dashed curve is shown for comparison: it tracks the similarity of the two series in a pair rather than the multifractality of their coupling, and its small, monotonically shrinking separation from $\lambda(q)$ reflects TASI's wider $h(q)$.}
   \label{fig:lambda}
 \end{figure}

The coupling carries a single scaling exponent. For every pair, $F^{q}_{xy}(s)$ retains a consistent
sign across the whole fitted range, so that the scaling relation (\ref{eq:lambda}) is satisfied and
$\lambda(q)$ is well defined. The resulting exponents are constant in $q$ to well within their own
uncertainty: over $q\in[0.5,4]$ the range $\Delta\lambda$ is $0.014$ (SAT--TSA), $0.023$ (SAT--TASI)
and $0.013$ (TSA--TASI), against a bootstrap standard error on a single $\lambda(q)$ of $0.02$--$0.04$
(Fig.~\ref{fig:lambda}, Table~\ref{tab:cross}). By the criterion of~\cite{Oswiecimka2014} the
cross-correlations are therefore described by a single cross-scaling exponent and carry no
multifractal structure of their own. This is the expected outcome given the univariate results of
Section~\ref{sec:origin}: no component series is multifractal within the resolution of the record, so
there is no multifractality available to propagate into the coupling.

The comparison with $[h_x(q)+h_y(q)]/2$ answers a different question---how alike the members of each
pair are in their scaling, rather than whether their coupling is
multifractal~\cite{Oswiecimka2014}---and it separates the pairs in a way that the widths do not. For
SAT--TASI the two curves coincide to within $0.009$ at every moment, so the gradient and the southern
regional index scale almost identically. The deviation is larger for the two pairs involving TSA,
reaching $0.032$ for TSA--TASI and $0.045$ for SAT--TSA, and in the latter case the mean of the
individual exponents falls outside the bootstrap band of $\lambda(q)$ over the lower half of the
moment range. The deviations are positive throughout, $\lambda(q)$ exceeding the mean of the component
exponents. This ordering follows directly from the shape of the individual $h(q)$ curves: SAT and TASI
have nearly flat exponents over the primary moment range and remain alike at every moment, whereas the
exponent of TSA increases with $q$ (Fig.~\ref{fig:h_tau_f}(a)), so that its mean with either partner
drifts away from the cross-scaling exponent as $q$ grows. The anomalous $q$-dependence identified in
Section~\ref{sec:origin} thus leaves a visible trace in the bivariate analysis as well, by a route
that does not involve $\Delta h$ at all.

The $q$-dependence of the coupling strength itself is shown in Fig.~\ref{fig:rhodcca}(b). For all
three pairs $\rho(q,s)$ varies smoothly and monotonically with $q$: at $s=41$~weeks the SAT--TASI
coefficient runs from $\approx-0.76$ at $q=0.5$ to $-0.67$ at $q=4$ and the TSA--TASI coefficient from
$-0.71$ to $-0.56$, so the anti-correlation is somewhat weaker among the largest joint excursions than
among the smallest, while the SAT--TSA coupling stays close to $+0.9$ throughout. The full $(q,s)$
planes are given in Supplementary~S4. This dependence is a property of the coupling's amplitude, not
of its scaling: a monotonic $\rho(q,s)$ is compatible with a single cross-scaling exponent, and the
multifractality of the coupling is diagnosed from $\lambda(q)$ rather than from
$\rho(q,s)$~\cite{Oswiecimka2014,Kwapien2015}.

\begin{table}[htbp]
\centering
\small
\setlength{\tabcolsep}{5pt}
\begin{tabular}{lcccccc}
\toprule
 & \multicolumn{3}{c}{$\rho_{\mathrm{DCCA}}(s)$} & \multicolumn{3}{c}{cross-scaling} \\
\cmidrule(lr){2-4}\cmidrule(lr){5-7}
Pair & $s=24$ & $s=44$ & $s=76$ & $\Delta\lambda$ & $\max_q|\lambda-\overline{h}|$ & scales signif. \\
\midrule
SAT--TSA  & $+0.887$ & $+0.915$ & $+0.945$ & $0.014$ & $0.045$ & $100\%$ \\
SAT--TASI & $-0.736$ & $-0.720$ & $-0.764$ & $0.023$ & $0.009$ & $100\%$ \\
TSA--TASI & $-0.625$ & $-0.617$ & $-0.693$ & $0.013$ & $0.032$ & $100\%$ \\
\bottomrule
\end{tabular}
\caption{Pairwise coupling of the Atlantic indices over the primary fitting range
$s\in[18,100]$ weeks, second-order detrending. The detrended cross-correlation coefficient
$\rho_{\mathrm{DCCA}}(s)$ is quoted at three representative scales spanning the range; every scale of
every pair lies outside the $95\%$ band of $100$ independent IAAFT surrogate pairs
($p=0.010$ throughout, the floor attainable with this ensemble size), as recorded in the last column.
$\Delta\lambda=\max_q\lambda(q)-\min_q\lambda(q)$ over $q\in[0.5,4]$ measures the multifractality of
the coupling itself and is to be compared with the bootstrap standard error on a single $\lambda(q)$,
$0.02$--$0.04$: all three pairs are consistent with a single cross-scaling exponent. The penultimate
column gives the largest departure of $\lambda(q)$ from the mean of the individual generalized Hurst
exponents, $[h_x(q)+h_y(q)]/2$, which measures how alike the members of each pair are rather than the
multifractality of their coupling~\cite{Oswiecimka2014}; the two pairs involving TSA, whose $h(q)$
increases with $q$, depart the most.}
\label{tab:cross}
\end{table}

%

\subsection{Association with ENSO: a lagged cross-correlation test}
\label{sec:oni}

Because the interhemispheric gradient of tropical Atlantic sea-surface
temperature has been associated with the El Ni\~no--Southern Oscillation, we
tested directly whether TASI is coupled to the Oceanic Ni\~no Index (ONI). Both
series were taken at monthly resolution over 1981--2025 ($530$ months; the weekly
TASI was aggregated to monthly means), and we computed the detrended
cross-correlation coefficient $\rho_{\mathrm{DCCA}}$ at a twelve-month scale as a
function of the lag $\tau$ between them ($\tau>0$: ONI leading). Crucially,
significance was assessed with a null that accounts for the search over lags: for
each of $200$ pairs of \emph{independent} IAAFT surrogates---which preserve the
autocorrelation of each series but destroy any mutual coupling---we recorded the
maximum $|\rho_{\mathrm{DCCA}}|$ over all lags, and compared the observed maximum
against this distribution (a family-wise test).

\begin{figure}[t]
   \centering
   \includegraphics[width=0.6\textwidth]{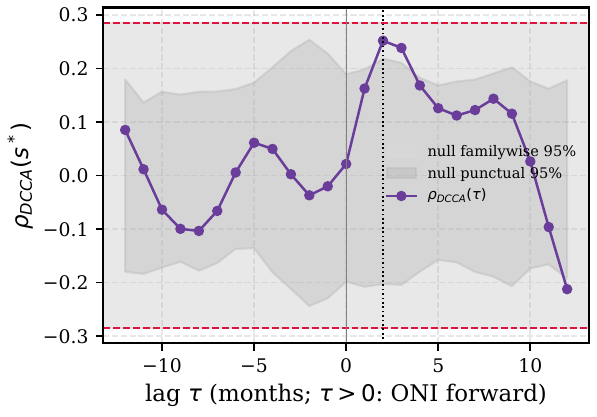}
   \caption{Lagged detrended cross-correlation $\rho_{\mathrm{DCCA}}(\tau)$
   between ONI and TASI at the twelve-month scale ($\tau>0$: ONI leading). Grey
   band: point-wise $95\%$ envelope of independent IAAFT surrogate pairs; dashed
   lines: the family-wise $5\%$ threshold ($\pm0.285$) that accounts for the
   search over lags. The peak at a two-month lead exceeds the point-wise band but
   not the family-wise threshold.}
   \label{fig:lag}
 \end{figure}

The coupling is strongest at a lead of two to three months
($\rho_{\mathrm{DCCA}}\approx0.25$ at $\tau=+2$), in the physically expected
direction of ENSO preceding the Atlantic response, and this peak does exceed the
point-wise $95\%$ surrogate band (Fig.~\ref{fig:lag}). A point-wise reading would
therefore flag it as significant. It does not, however, exceed the family-wise
threshold ($\pm0.285$; $p=0.16$): a coupling of this magnitude arises in roughly
one pair in six of independent, equally persistent series purely from their
autocorrelation. The estimated lead is moreover unstable, shifting from
$\tau=+2$ at the twelve-month scale to $\tau=+7$ when the coefficient is averaged
over scales. Once the strong persistence of both indices and the multiplicity of
tested lags are taken into account, we thus find no statistically significant
coupling between ENSO and the Atlantic gradient. We accordingly make no causal
claim: the suggestive lead--lag signal is consistent with the coincidental
alignment of two highly autocorrelated series, and directional diagnostics that
presuppose a coupling (Granger causality, transfer entropy) are not warranted on
this evidence.

\section{Conclusions}
\label{sec:conclusions}

We have reassessed the scaling properties of three tropical Atlantic SST indices, with particular
attention to the conditions under which a multifractal width can be estimated at all from a record of
this length. Three choices proved decisive, and none of them is a matter of numerical detail. The
first is the fitting range: exponents are estimated over $s\in[18,100]$ weeks, the lower bound being
the smallest scale at which a second-order detrending is well determined
($s_{\min}\gtrsim6(m+1)$ for order $m$) and the upper bound stopping short of the change of slope that
the fluctuation functions display beyond $\approx100$ weeks. The second is the moment range,
restricted to $|q|\le4$, beyond which the fluctuation function is governed by a handful of extreme
segments. The third is the comparison of every width against ensembles of surrogates rather than
against a single realization.

Once these are in place, the conclusion is uniform across the three indices: none of them exhibits
multifractality that survives comparison with its surrogate ensembles. The widths of SAT and TASI
($\Delta h=0.077$ and $0.054$) lie inside both nulls. The width of TSA ($\Delta h=0.217$) exceeds
both, but is excluded on structural rather than statistical grounds: its generalized Hurst exponent
increases with $q$, and since $f(\alpha)=1+q^{2}h'(q)$, an increasing $h(q)$ forces $f(\alpha)>1$, a
value no one-dimensional record can attain. A multiplicative cascade generates $h'(q)<0$; what is
measured here has the opposite sign, and is the signature of a single power law fitted across a change
of scaling regime rather than of a cascade. Within the resolution afforded by the record, the two
regional indices and the interhemispheric gradient are alike monofractal on the scales over which a
power law holds.

This revises our own earlier estimates as well as those in the literature. The apparent multifractal
width is not identifiable at this record length, and the demonstration is direct: the ordering of the
three indices by $\Delta h$ reverses with the fitting range, with the moment range and with the
detrending order, so that each of the three is the widest of the three under some combination of
choices that is defensible on its own terms. Within every configuration the bootstrap confidence
intervals of the three indices overlap. With $N\approx2300$ weekly observations and about
three-quarters of a decade of usable scales, $\Delta h$ is a quantity about which this record does not
speak, and the dynamical distinction between regional variability and the interhemispheric
gradient---which earlier estimates, including our own, located in the multifractal width---is not
supported.

What the record does establish is the persistence. The generalized Hurst exponent $h(2)$ lies between
$1.19$ and $1.42$ and exceeds unity for every index at every detrending order tested, by at least
seven bootstrap standard errors, so the three indices are strongly persistent and locally
non-stationary on the analysed scales. Its magnitude depends on the detrending order, rising by
between $0.12$ and $0.22$ from DFA-1 to DFA-3 even with $s_{\min}$ raised in step with the order, and
should be quoted with the order attached; the qualitative diagnosis does not. The moving-window
analysis adds that this persistence varies appreciably over the record, and robustly so across window
lengths, revealing epochs of stronger and weaker persistence shared by the three series, while the
local multifractal width never exceeds its window-specific surrogate level in any index or any epoch.
The time dependence that is robustly established is that of the persistence, not of any multifractal
property.

The cross-correlation analysis is unaffected by the reclassification, since it never depended on the
individual widths. The two regional indices co-vary positively and increasingly towards longer scales
($\rho_{\mathrm{DCCA}}$ rising from $0.88$ to $0.97$), whereas the interhemispheric gradient is
anticorrelated with both ($\rho_{\mathrm{DCCA}}\approx-0.7$), as expected of a contrast index. In every
pair the cross-scaling exponent is constant in $q$ to within its own uncertainty and coincides with the
mean of the individual Hurst exponents, so the coupling carries a single scaling exponent and no
multifractal structure of its own. A direct test of the association with ENSO reinforces the need for
caution of a different kind: the detrended cross-correlation between the ONI and the Atlantic gradient
peaks in the physically expected direction, with ENSO leading by two to three months, and would be
judged significant by a point-wise criterion; but it does not survive a surrogate null that accounts
for the search over lags, and the estimated lead is unstable across scales. We therefore report no
significant ENSO--Atlantic coupling, and attribute the suggestive lead--lag signal to the shared
persistence of the two indices rather than to a genuine dynamical link.

Taken together, these results describe the tropical Atlantic surface temperature indices as strongly
persistent, locally non-stationary and monofractal within the resolution of the record, strongly
coupled but with a single cross-scaling exponent, and not demonstrably linked to ENSO once the
multiplicity of tested lags is accounted for. The methodological point generalizes beyond these
records. A reported multifractal width should not be interpreted before three conditions have been
checked: that the fitted range contains no change of scaling regime, since a single exponent fitted
across two of them returns a $q$-dependence that belongs to neither; that the sign of $h'(q)$ is
negative throughout, since $f(\alpha)=1+q^{2}h'(q)$ makes an increasing $h(q)$ inconsistent with any
admissible singularity spectrum; and that the width is stable under the discretionary choices of
moment range and detrending order. These are inexpensive checks, and each of them is decisive here.
Where they fail, a high coefficient of determination offers no protection---all three indices scale
with $R^{2}\ge0.975$ over the fitted range while yielding widths that reverse their ordering under
every choice examined.

\section*{Data and Code Availability}
The SAT and TSA indices are available from the NOAA Physical Sciences Laboratory. The ONI data was retrieved from the NOAA Climate Prediction Center. All core numerical computations---including the global and moving-window MFDFA, the surrogate ensembles, and the sign-preserving cross-correlation (MFCCA) and lagged cross-correlation analyses---were executed using the open-source \texttt{mf-toolkit} Python library, which leverages the algorithmic structures validated in \cite{mendez2026mf}.

\bibliographystyle{unsrt}
\bibliography{SST_refs}

 \section*{Supplementary Material}

section*{S1. Scaling of the fluctuation function and quality of the fits}

\section*{S1. Scaling of the fluctuation function and the fitted range}

The generalized Hurst exponents reported in the main text are the slopes of least-squares fits of
$\log_{10}F_q(s)$ against $\log_{10}s$ over $s\in[18,100]$ weeks. Their reliability rests on the
fluctuation functions actually scaling as power laws over that range, and on the range being chosen
before the exponents are inspected rather than after. We document both here.

Figure~S1(a--c) shows $F_q(s)$ for SAT, TSA and TASI at representative moments spanning $|q|\le4$,
computed over the full accessible interval $s\in[18,230]$ weeks, with the fitted range and its
complement distinguished. The lower bound of the fit is set by the detrending order: a polynomial of
order $m$ is not determined within a segment appreciably shorter than $6(m+1)$ points, so that
$s_{\min}=18$ for the second-order detrending adopted here. Below that bound the polynomial absorbs a
progressively larger share of the variance within each segment, and the resulting values of $F_q(s)$
are not estimates of a fluctuation. The upper bound is set by the data. Over the first
three-quarters of a decade the curves are linear on log--log axes for every index and every moment;
beyond $s\approx100$ weeks the slope decreases visibly, and the extrapolation of the fitted line
(dotted) runs above the measured points. The change of slope is present in all three indices and at
every moment shown.

This is why the fit is not carried across the whole interval. A single power law fitted over
$[18,230]$ does not estimate the exponent of either regime: it returns a compromise between the two,
weighted by the number of points each contributes and by how the two slopes are arranged in $q$. That
compromise is itself $q$-dependent, so it enters $h(q)$ as a spurious tilt whose sign and magnitude
differ between records. Section~\ref{sec:origin} quantifies the consequence: extending the fit past
$100$ weeks raises the apparent width of TASI from $0.054$ to $0.239$ and lowers that of TSA from
$0.217$ to $0.010$, reversing which index appears the most multifractal of the three.

It should be emphasised that the quality of the fit gives no warning of this. Figure~S1(d) reports
the coefficient of determination moment by moment. Over the primary range $R^{2}\ge0.975$ for every
index and every moment in $|q|\le4$, with minima of $0.975$ (SAT), $0.991$ (TSA) and $0.992$ (TASI). A high $R^{2}$ over
a range that contains a change of regime therefore certifies that the dominant stretch scales, not
that the range as a whole does. This is a general caution and not a feature of these particular
records: the coefficient of determination is not a diagnostic for the presence of a crossover, and
the fitted range must be established from the shape of $F_q(s)$ and from the constraints of the
detrending order, as it has been here.
\begin{figure}[h!]
    \centering
    \begin{subfigure}[b]{0.48\linewidth}
        \centering
        \includegraphics[width=\linewidth]{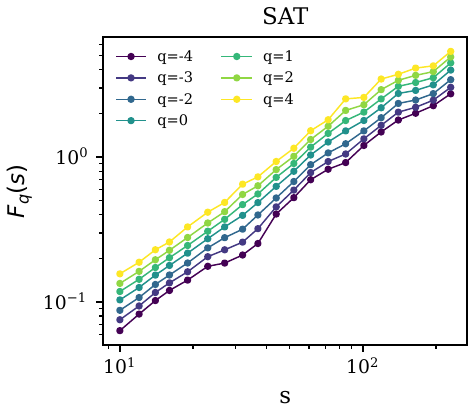}
        \caption{}
    \end{subfigure}
    \hfill
    \begin{subfigure}[b]{0.48\linewidth}
        \centering
        \includegraphics[width=\linewidth]{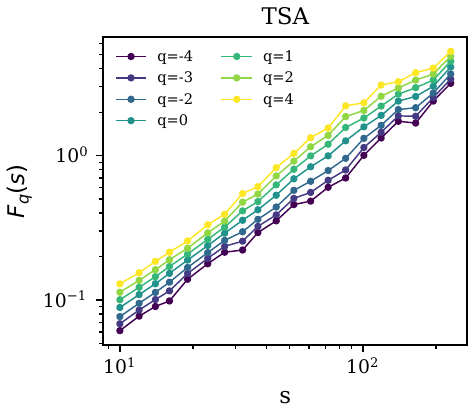}
        \caption{}
    \end{subfigure}

    \vspace{4pt}

    \begin{subfigure}[b]{0.48\linewidth}
        \centering
        \includegraphics[width=\linewidth]{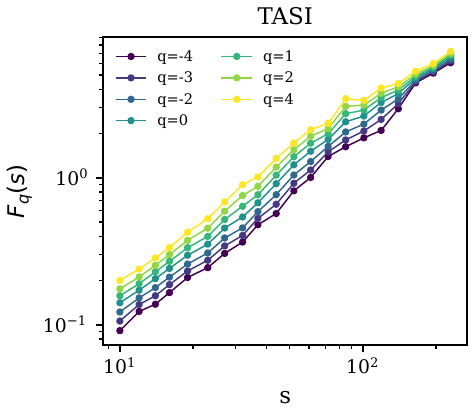}
        \caption{}
    \end{subfigure}
    \hfill
    \begin{subfigure}[b]{0.48\linewidth}
        \centering
        \includegraphics[width=\linewidth]{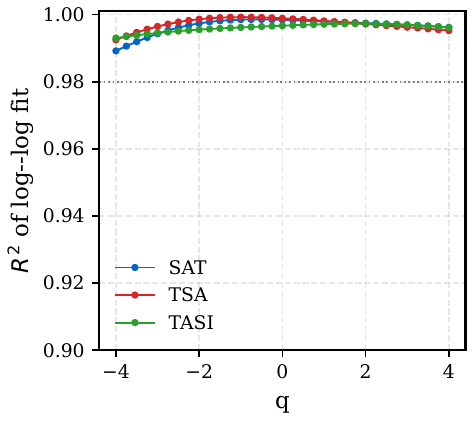}
        \caption{}
    \end{subfigure}

    \caption*{\textbf{Figure S1.} Scaling of the fluctuation function, second-order detrending.
    (a--c)~$F_q(s)$ on log--log axes for SAT, TSA and TASI at representative moments spanning the
    primary range $|q|\le4$, shown over the full accessible interval $s\in[18,230]$ weeks. The fitted range is $s\in[18,100]$. The curves are linear over the fitted range for every index and every moment; beyond $s\approx100$ weeks the slope decreases and the measured
    points fall below the extrapolation. (d)~Coefficient of determination of the log--log fit over the fitted range, as a function of $q$; $R^{2}\ge0.975$ for every index and every moment of the primary range.}
\end{figure}

\section*{S2. Distribution of the increments}

The interpretation of Section~\ref{sec:origin}---that the residual multifractal width of TASI originates in nonlinear temporal correlations rather than in a broad distribution of fluctuations---rests in part on the shape of the increment distribution, which we document here.

Figure~S3 shows the complementary cumulative distribution function of the standardised absolute weekly increments, $|\Delta x|/\sigma$, for the three
indices, together with that of a standard Gaussian variable ($\mathrm{erfc}(z/\sqrt{2})$).

The three indices follow the Gaussian reference over the bulk of the distribution and depart from it only beyond $\approx3\sigma$, where a
modest excess appears. This is consistent with the measured excess kurtosis of $0.5$--$0.9$: the increments are mildly leptokurtic, but they are not
heavy-tailed in the sense relevant here. 

Two points follow, and neither requires the increments to be exactly Gaussian. First, the surrogate test does not depend on the shape of the distribution at all: the IAAFT algorithm preserves the amplitude distribution of the original record by construction, so any tail excess---however large---is reproduced in the null ensemble, and a measured width that exceeds that ensemble cannot be ascribed to it. Second, the magnitude of the departure matters for the residual interpretation. Because a broad distribution can widen a singularity spectrum only in the presence of temporal correlations~\cite{Kwapien2023,Rak2025}, the relevant question is not whether the increments are Gaussian but whether their
tails are heavy enough to contribute appreciably once correlations are present. At an excess kurtosis below unity they are not: these records sit far from the regime---q-Gaussian with $q$ well above one, or Lévy-stable---in which the distributional contribution to the spectral width becomes significant.

The same observation constrains the moment range. The sensitivity of $F_q(s)$ to outliers at large $|q|$, noted in Section~\ref{sec:origin}, is a sensitivity to extreme segments of the \emph{fluctuation function}, not to extreme increments: with increments this close to Gaussian, the dominance of a handful of segments at $|q|\gtrsim5$ (Fig.~\ref{fig:h_tau_f}(b)) reflects the finite number of segments available at the largest scales rather than a genuinely heavy-tailed generating process. This is a further reason to restrict the analysis to $|q|\le4$.

\begin{figure}[h!]
    \centering
    \includegraphics[width=.6\linewidth]{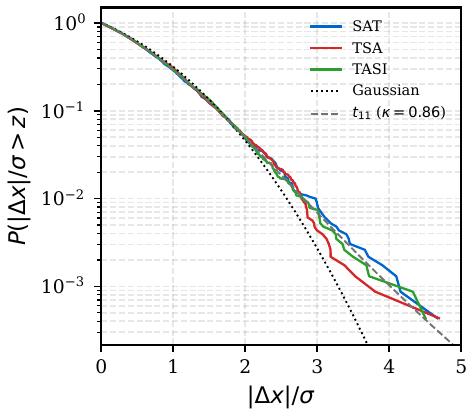}
    \caption*{\textbf{Figure S3.} Complementary cumulative distribution function     of the standardised absolute weekly increments $|\Delta x|/\sigma$ for SAT, TSA and TASI, compared with a standard Gaussian variable (dotted). The     increments are mildly leptokurtic (excess kurtosis $0.5$--$0.9$) and far     from the heavy-tailed regime in which the distribution of fluctuations     contributes appreciably to the width of the singularity spectrum.}
    \label{fig:ccdf}
\end{figure}

\section*{S3. The $q$-dependent cross-correlation coefficient over the full
$(q,s)$ plane}

The coupling analysis of Section~\ref{sec:entropy_and_mfdcca} uses two sections
through a single quantity. The detrended cross-correlation coefficient
$\rho_{\mathrm{DCCA}}(s)$ of Eq.~(\ref{eq:rhodcca}) is the $q=2$ section of the
$q$-dependent coefficient $\rho(q,s)$ of Eq.~(\ref{eq:rhoq}), and
Fig.~\ref{fig:rhodcca}(b) is the section at the single scale $s=52$ weeks. Both
are reported in the main text because they answer the questions the analysis
poses: how the coupling strength varies with time scale, and how it varies with
the amplitude of the joint fluctuations. This supplement gives the full plane
from which both are drawn, so that neither section is taken on trust.

Figure~S4 shows $\rho(q,s)$ for the three pairs over the whole domain of the
analysis: $q\in[0.5,4]$, the range in which the Cauchy--Schwarz bound
$-1\le\rho(q,s)\le1$ holds (Section~\ref{sec:entropy_and_mfdcca}), and
$s\in[10,230]$ weeks, the fitted scaling range. Colour encodes the coefficient
on a symmetric scale; dots mark the cells whose coefficient does not reach
significance at the $5\%$ level against the null of 100 pairs of independent
IAAFT surrogates, which preserve the autocorrelation of each series while
destroying their mutual coupling. The $q=2$ row of each panel is the curve of
Fig.~\ref{fig:rhodcca}(a) and the $s=52$ column is the curve of
Fig.~\ref{fig:rhodcca}(b).

The planes are almost entirely significant: of the $720$ cells resolved for each
pair---$20$ scales by $36$ moments---none fail for SAT--TSA, one for SAT--TASI
and four for TSA--TASI, all of them at the longest scale, $s=230$ weeks. The
coupling of the two regional indices is strong and positive throughout,
$\rho(q,s)$ lying between $+0.82$ and $+0.99$, and it shows no feature at the
longest scales: between $s=140$ and $s=230$ weeks it changes by less than $0.01$
at every moment order.

The two pairs involving the gradient behave differently, and the plane shows
that the weakening reported in Section~\ref{sec:entropy_and_mfdcca} is not
confined to the $q=2$ section from which it was described. Between $s=140$ and
$s=230$ weeks the SAT--TASI coefficient weakens by $0.30$ at $q=0.5$ and by
$0.20$ at $q=4$, and the TSA--TASI coefficient by $0.38$ and $0.21$
respectively; the effect is present at every moment order and is somewhat
stronger for the smaller joint fluctuations than for the larger ones. That the
SAT--TSA plane is flat over exactly the same scales, at every $q$, is what
rules out a generic artefact of the reduced segment count there: the decoupling
belongs to the gradient's long-scale relation with the regional indices, not to
the edge of the fitted range.

\begin{figure}[htbp]
    \centering
    \begin{subfigure}[b]{0.32\textwidth}
        \includegraphics[width=\textwidth]{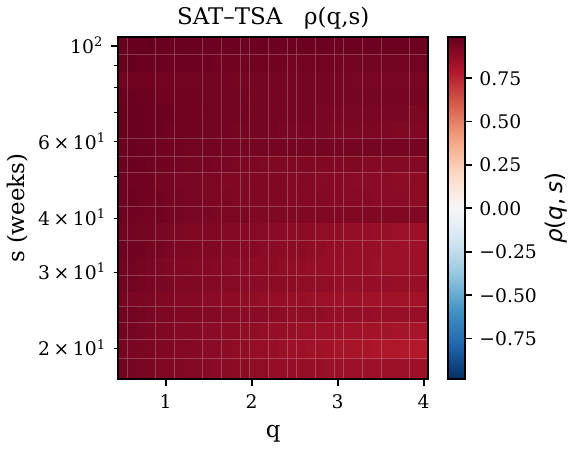}
        \caption{}
    \end{subfigure}
    \hfill
    \begin{subfigure}[b]{0.32\textwidth}
        \includegraphics[width=\textwidth]{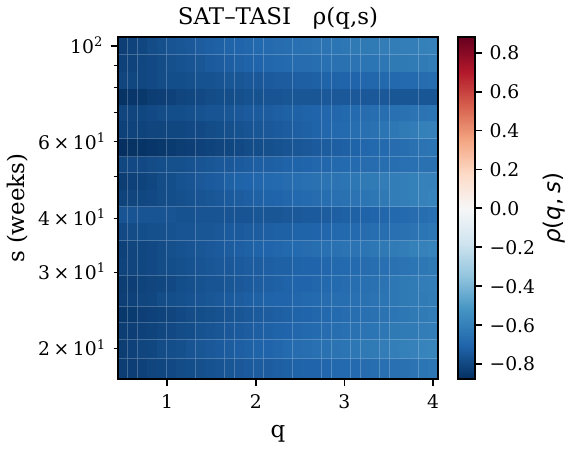}
        \caption{}
    \end{subfigure}
    \hfill
    \begin{subfigure}[b]{0.32\textwidth}
        \includegraphics[width=\textwidth]{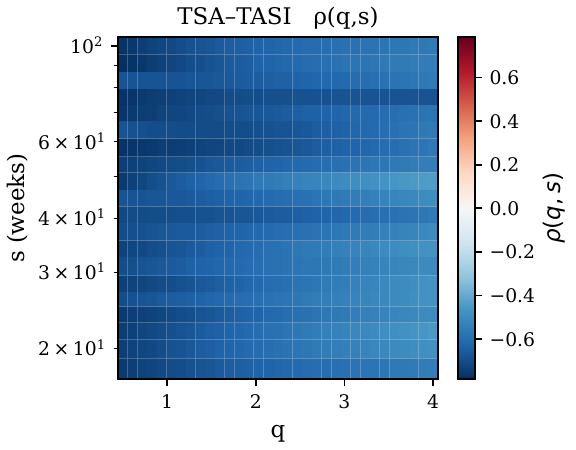}
        \caption{}
    \end{subfigure}
    \caption*{\textbf{Figure S4.} The $q$-dependent detrended cross-correlation
    coefficient $\rho(q,s)$, Eq.~(\ref{eq:rhoq}), over the full $(q,s)$ plane for
    (a)~SAT--TSA, (b)~SAT--TASI and (c)~TSA--TASI, with second-order detrending.
    The vertical axis is logarithmic in the scale. Dots mark cells that do not
    reach significance at the $5\%$ level against 100 pairs of independent IAAFT
    surrogates.}
    \label{fig:S4}
\end{figure}

\end{document}